\documentclass[letterpaper,twocolumn,10pt]{article}
\usepackage{usenix,epsfig,endnotes}

\usepackage{graphicx}
\usepackage{amsmath}
\usepackage{amssymb}
\usepackage{amsfonts}
\usepackage{booktabs}
\usepackage{times}
\usepackage{microtype}
\usepackage{epsfig}
\usepackage{caption}
\usepackage{float}
\usepackage{placeins}
\usepackage{color, colortbl}
\usepackage{stfloats}
\usepackage{enumitem}
\usepackage{tabularx}
\usepackage{xstring}
\usepackage{multirow}
\usepackage{xspace}
\usepackage{url}
\usepackage{subcaption}
\usepackage{xcolor}
\usepackage[hang,flushmargin]{footmisc}
\usepackage{soul}
\usepackage{amsthm}
\usepackage[ruled,vlined,linesnumbered]{algorithm2e}
\usepackage[noend]{algpseudocode}
\usepackage{bbold}
\usepackage{tcolorbox}

\usepackage{makecell}
\usepackage{tikz}
\usepackage{array}
\usepackage{diagbox}
\usepackage{hyperref}
\usepackage{tablefootnote}
\usepackage{todonotes}
\usepackage{booktabs}
\usepackage{adjustbox}

\usepackage{{multirow}}

\usepackage[utf8]{inputenc}
\usepackage{amssymb}
\usepackage[normalem]{ulem}

\usepackage{graphicx}

\usepackage{placeins}

\renewcommand{\arraystretch}{1.5}

\usepackage{tikz}

\usepackage{booktabs}
\usepackage{multirow}
\usepackage{siunitx}
\usepackage{todonotes}
\usepackage{comment}
\usepackage{threeparttable}

\newcommand{\R}[1]{{%
    \textbf{%
        \ifstrequal{#1}{1}{\textcolor{red}{R#1}}{%
        \ifstrequal{#1}{2}{\textcolor{blue}{R#1}}{%
        \ifstrequal{#1}{3}{\textcolor{magenta}{R#1}}{%
        \ifstrequal{#1}{4}{\textcolor{teal}{R#1}}{%
                           \textcolor{cyan}{R#1}%
        }}}}%
    }%
}}

\newcommand{\PP}[1]{
\vspace{2px}
\noindent{\bf {#1.}}}

\theoremstyle{definition}

\newcolumntype{Y}{>{\raggedright\arraybackslash}X}

\newif\ifarxiv 

\definecolor{softblue}{RGB}{198, 218, 236}
\definecolor{softorange}{RGB}{246, 216, 188}
\definecolor{softgreen}{RGB}{204, 232, 207}
\definecolor{softpurple}{RGB}{220, 204, 232}
\definecolor{softyellow}{RGB}{248, 238, 190}

\definecolor{softred}{RGB}{242, 200, 200}
\definecolor{softrose}{RGB}{238, 210, 220}
\definecolor{softsalmon}{RGB}{244, 204, 195}

\definecolor{softgrey}{RGB}{220, 220, 220}

\usepackage{xurl}
\makeatletter
\let\ps@plain\ps@empty
\makeatother

\begin{document}

\pagestyle{empty}
\pagenumbering{gobble}

\date{}

\title{\Large \bf What the Eyes See, the LLMs Miss: Exploiting Human Perception for Adversarial Text Attacks}

\author{
{\rm Qin Yang$^{1}$\thanks{Both authors contributed equally to this work.}, Lu Malloy$^{2}$\footnotemark[1], Joshua Lee$^{3}$, Xiaohan Chang$^{1}$, Meisam Mohammady$^{4}$}\\
{\rm Doowon Kim$^{2}$, Yuan Hong$^{1}$}\\[0.4em]
$^{1}$\textit{University of Connecticut},
$^{2}$\textit{University of Tennessee},\\
$^{3}$\textit{University of California, Santa Barbara},
$^{4}$\textit{Iowa State University}
}

\maketitle

\thispagestyle{empty}
\pagestyle{empty}

\begin{abstract}
Large language model (LLM)-powered content moderation systems are a critical defense against harmful online content. However, they operate primarily on tokenized text and often overlook visual cues that humans naturally use when interpreting content. We show that this limitation creates a fundamental vulnerability: content readily recognized as harmful by humans can evade automated moderation. To systematically study this problem, we introduce Human-Perceptible Adversarial Attacks (HPAA), which embed harmful expressions into otherwise benign text using visually salient typographic manipulations. HPAA strategically combines features such as spacing, emphasis, and spatial arrangement to preserve human recognition while reducing machine detectability. Operating in a black-box setting with a small query budget, the attack automatically generates evasive content without model access or gradient information. We evaluate HPAA on multiple datasets and thirteen widely deployed moderation systems, including commercial APIs and state-of-the-art open-source guardrails. With only three detector queries, generated attacks achieve over 86\% human recognition while keeping detection rates below 1\% across evaluated systems. We further identify the typographic factors driving successful evasion, analyze why current moderation architectures fail to capture these signals, and discuss practical defenses. Our findings reveal a fundamental blind spot in current LLM-based moderation systems and motivate moderation approaches that better align with human perceptual understanding.\footnote{Artifacts are available at \href{https://github.com/datasec-lab/hpaa}{\textcolor{purple}{https://github.com/datasec-lab/hpaa}}. This is the full version of the paper to appear at USENIX Security 2026.} \textcolor{red}{Disclaimer: This paper includes examples of harmful, hateful, or abusive language for research purposes. Reader discretion is advised.}
\end{abstract}

\section{Introduction}
\label{sec:introduction}

Large language models (LLMs) have become a core component of modern content-moderation pipelines~\cite{franco2025integrating}, enabling large-scale detection and filtering of harmful, hateful, misleading, and policy-violating content~\cite{chen2025comprehensive,aldahoul2024advancing}. Widely deployed across social media platforms and online communities, these systems often serve as the first line of defense before content reaches end users. Consequently, failures in automated moderation can directly expose users to harmful content at scale, making moderation robustness an increasingly important security concern~\cite{bender2021dangers,weidinger2021ethical,wu2019misinformation}.

Existing attacks against language models primarily focus on manipulating model behavior. Adversarial text attacks use lexical, syntactic, or semantic perturbations to induce misclassification while preserving meaning~\cite{li2018textbugger,ebrahimi2018hotflip,gao2018black,jin2020bert,garg2020bae,feng2018pathologies,pruthi2019combating,dyrmishi2023humans}. Jailbreaking and prompt-injection attacks circumvent safety mechanisms by manipulating prompts or model instructions~\cite{geng2025safety,greshake2023youvesignedforcompromising,Yi_2025}. Toxicity-evasion attacks seek to bypass moderation systems through rephrasing, masking, or stylization of harmful content~\cite{vidgen2019challenges,mathew2021hatex,shen2025hatebench,zhu2025taebench,koh2024can}. Despite their differences, these attacks are evaluated primarily through model behavior and implicitly assume that humans and models perceive content in broadly similar ways.

In this paper, we investigate a fundamentally different vulnerability arising from a mismatch between human perception and machine processing. Human readers rely not only on lexical content but also on visual cues such as spacing, emphasis, and spatial arrangement when interpreting messages. Modern moderation systems, however, operate primarily on tokenized textual representations and often disregard these typographic signals. This discrepancy raises a fundamental question: \emph{Can harmful content remain readily recognizable to humans while evading automated moderation systems?}

We answer this question affirmatively. We introduce \textbf{Human-Perceptible Adversarial Attacks (HPAA)}, a new class of attacks that embed harmful content into otherwise benign text through human-interpretable typographic patterns. Unlike conventional adversarial attacks that seek to confuse either humans or models, HPAA exploits the gap between them: harmful content remains perceptible and meaningful to human recipients while becoming difficult for automated moderation systems to detect.

\begin{figure*}
    \centering
    \includegraphics[width=0.9\linewidth]{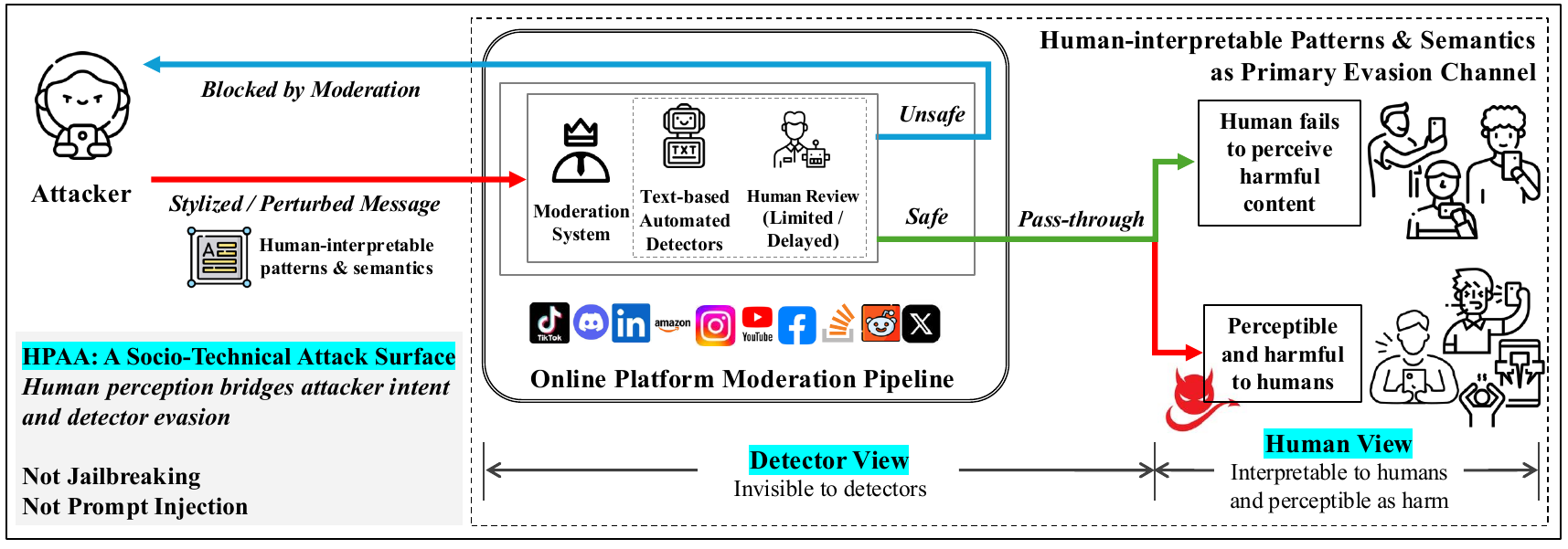}
    \caption{Overview of typographic Human-perceptible Adversarial Attacks (HPAA) against modern content moderation pipelines. Harmful content remains recognizable to human readers while evading automated moderation systems.}
    \label{fig.overview}
\end{figure*}

Figure~\ref{fig.overview} illustrates the core intuition. In modern moderation pipelines, content is typically screened by automated detectors before being delivered to human users. Under this architecture, an attacker need not deceive both humans and machines. Instead, the attacker only needs to evade the automated moderator while preserving the harmful message for the eventual recipient. HPAA exposes a previously underexplored \emph{socio-technical attack surface} where human perception effectively bridges attacker intent and moderation evasion.

Realizing such attacks introduces several unique challenges that are absent from conventional moderation-evasion settings. Unlike prior work that evaluates success solely through automated classifiers, HPAA requires simultaneously optimizing two competing objectives: reducing detectability by moderation systems while preserving human recognition of harmful content. Demonstrating attack effectiveness therefore requires rigorous human-centered evaluation methodologies that explicitly account for human perception in addition to moderation outcomes.

To systematically study this vulnerability, we develop a practical HPAA framework that leverages simple typographic cues, including spacing, visual emphasis, and spatial arrangement, to encode harmful content within otherwise benign text. The attack operates in realistic black-box settings, requires no access to model internals or gradients, and can be generated with a relatively small query budget. We evaluate HPAA across multiple datasets and thirteen widely deployed moderation systems, including both commercial APIs and state-of-the-art open-source guardrails. Our results reveal a striking and consistent gap between human and machine perception: content that remains highly recognizable to human readers can achieve extremely low detection rates across modern moderation systems. These findings expose a fundamental blind spot in current LLM-based moderation architectures and motivate more robust moderation approaches that more effectively incorporate human-perceptible signals.

\section{Preliminaries}
\label{sec:preliminaries}

\subsection{Content Moderation Pipelines}
\label{sec:prelim-moderation}

Modern online platforms commonly employ tiered moderation pipelines to screen user-generated content before it reaches end users. Submitted content (e.g., posts, comments, or messages) is first processed by an \emph{automated moderator}, which classifies content as \emph{safe} or \emph{unsafe}. Content associated with low confidence or ambiguous decisions may be escalated for human review. Automated moderators are typically implemented using toxicity classifiers, including traditional \textit{BERT}-based models~\cite{unitary_toxic_bert, unitary_multilingual_xlm_roberta, textdetox_bert_multilingual_toxicity, snlp_roberta_toxicity_classifier,logacheva-etal-2022-paradetox,devlin2018bert,liu2019roberta,conneau-etal-2020-unsupervised} and modern \textit{LLM}-based systems~\cite{inan2023llama, perspectiveapi_website}.

Without loss of generality, we model the automated moderator as a toxicity detector $f(\cdot)$ that maps an input text to either a binary decision or a toxicity score. Our formulation is agnostic to the underlying architecture and applies to both LLM-based and conventional classifiers. Throughout this paper, we use the terms \emph{toxic}, \emph{harmful}, and \emph{unsafe} interchangeably to refer to content that violates platform policies, including hate speech, threats, harassment, sexual content, self-harm--related expressions, and other abusive language.\footnote{The terms `toxic,'' `harmful,'' and ``unsafe'' are used interchangeably.}

In practice, moderation systems operate as imperfect filters rather than oracle classifiers. Even state-of-the-art detectors exhibit nonzero false-positive and false-negative rates, requiring platform operators to balance safety, usability, and moderation cost. As a result, many platforms employ human moderators to review ambiguous, high-risk, or contested cases that detectors cannot confidently resolve. Nevertheless, as primary screening mechanism at scale, detectors must generalize across diverse content domains and attack strategies. This operational constraint creates opportunities for adversaries to exploit signals that remain salient to human readers but are weakly captured by automated moderation systems.

\subsection{Typographic Configuration Space}
\label{sec:prelim-notation}

Human readers do not rely solely on lexical content when interpreting text. Visual cues such as spacing, spatial arrangement, and visual emphasis also influence how information is perceived and reconstructed. Many of these cues are naturally exploited in everyday communication to draw attention, convey structure, or emphasize particular messages. In contrast, modern moderation systems primarily operate on tokenized textual representations and may only partially capture the information conveyed through visual presentation.

To systematically study the role of typography in moderation evasion, we organize typographic manipulations into three complementary dimensions: \emph{typographic granularity} ($L$), \emph{placement strategy} ($M$), and \emph{stylistic transformation} ($S$). Granularity determines how toxic content is decomposed prior to embedding, placement specifies how toxic fragments are arranged within surrounding benign text, and style controls the visual appearance of selected characters or tokens. Together, these dimensions define a structured space of human-perceptible typographic modifications.

Formally, we define the \emph{typographic configuration space} as $\mathcal{H} = \{\mathrm{M}, \mathrm{L}, \mathrm{S}\}$, where $\mathrm{M}$ denotes positional patterns, $\mathrm{L}$ specifies the granularity of toxic-span decomposition, and $\mathrm{S}$ represents stylistic transformations. A configuration $h \in \mathcal{H}$ determines how toxic content is embedded within benign text and presented to both human readers and automated moderation systems. The configuration space grows combinatorially with choices of $\mathrm{M}$, $\mathrm{L}$, and $\mathrm{S}$, enabling both restricted and expanded instantiations. In this work, we evaluate a representative subset of configurations motivated by empirical observations and practical deployment constraints. The framework itself is not tied to any specific configuration choices; additional typographic transformations can be incorporated within the same formulation. Details can be found in Section~\ref{sec:typographic-cues}.

\subsection{User Study and Ethics}
\label{sec:prelim-human}

Human evaluation is a central component of this work because HPAA is designed to exploit discrepancies between human perception and automated moderation. Consequently, attack success cannot be measured solely through moderation-evasion rates; it must also account for whether the embedded harmful content remains recognizable to human readers.

To assess human perceptibility, we conduct controlled user studies involving human participants. All study procedures were reviewed and approved by our Institutional Review Board (IRB). Participants were recruited through an online panel, provided informed consent, and were compensated according to platform guidelines. The study included content warnings, allowed withdrawal at any time, and collected no personally identifiable information. Detailed descriptions of the study protocol, participant demographics, perceptibility measures, and statistical analyses are provided in Section~\ref{sec:user_study}.

\section{Attack Design}
\label{sec:design}

\subsection{Problem Formulation}
\label{sec:problem}

We consider a black-box moderation setting in which an attacker can only query a content moderation system and observe its outputs. Given a toxic input, the attacker aims to generate a typographically modified sample that remains recognizable as harmful to human readers while avoiding detection by the moderation system. Unlike conventional adversarial text attacks, which primarily focus on inducing model misclassification, our attack aims to create a discrepancy between human perception and detector judgments. Specifically, we seek typographic configurations that preserve human recognition of harmful content while causing automated moderation systems to classify the content as benign. Given the dissemination and amplification mechanisms of online platforms, harmful content that remains recognizable to humans while bypassing automated moderation may propagate in a subtle and difficult-to-detect manner. Furthermore, because the ultimate recipients of moderated content are human users, analyses based solely on model behavior provide an incomplete characterization of real-world moderation risks.

To systematically characterize this human--LLMs perceptual mismatch and guide the design of HPAA, we investigate the following research questions:

\begin{itemize}[leftmargin=*, topsep=0pt, itemsep=0em]
    \item \textbf{RQ1:} To what extent can human readers recognize toxic content embedded within typographically manipulated text, and how does recognition vary across different typographic configurations?
    \item \textbf{RQ2:} To what extent do automated moderation systems detect typographically manipulated toxic content, and how does their performance differ from human recognition?
    \item \textbf{RQ3:} To what extent can the typographic cue configurations identified from human and moderation analyses enable effective moderation evasion in practical settings?
\end{itemize}

 \begin{figure*}[h]
    \centering
    \includegraphics[width=0.96\linewidth]{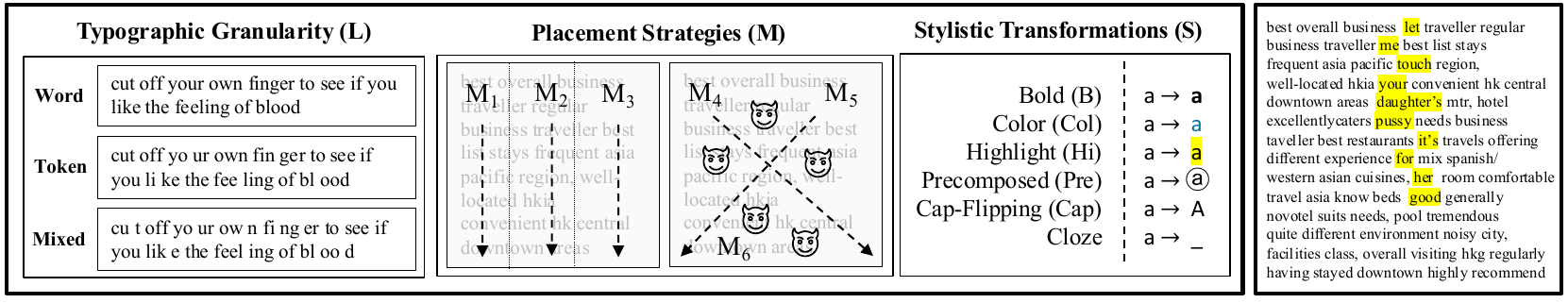}
    \caption{Typographic configuration space in HPAA and an illustrative example (derived from real-world toxic content).}
    \label{fig:pipeline}
\end{figure*}

\subsection{Threat Model}
\label{sec:threat}

\textbf{Attacker.}
We consider a non-privileged adversary who submits content through standard online platforms (e.g., social media or online forums) using ordinary user accounts.

\vspace{0.05in}

\noindent \textbf{Knowledge and Capabilities.}
The adversary has no access to the moderation system's architecture, parameters, training data, or moderation logic. The attacker interacts with the moderation pipeline in a purely black-box manner and can only observe moderation outcomes, such as safe/unsafe decisions and, when available, confidence scores. The attacker is restricted to content-level manipulations and cannot influence platform policies, moderation rules, or system deployment. The adversary may additionally obtain limited feedback on human perceptibility through ordinary user interactions or small-scale surveys. Such feedback is used only to estimate human readability and does not provide any information about the moderation system or its internal decision process.

\vspace{0.05in}

\noindent \textbf{Goal.}
The attacker's goal is to generate content that remains recognizable as harmful to human readers while receiving a benign moderation decision, thereby enabling the content to bypass automated moderation and reach end users.

\vspace{0.05in}

\noindent \textbf{Threat Surface.}
We consider the automated content moderation pipeline as the primary threat surface, including the text submission interface and the moderation mechanisms that determine whether content is released, blocked, or escalated for further review.

\subsection{Typographic Configuration Space}
\label{sec:typographic-cues}

The key idea behind HPAA is that humans do not rely solely on lexical content when interpreting text. Instead, visual cues such as spacing, spatial arrangement, and visual emphasis also influence how information is perceived and reconstructed. To systematically study how these cues affect both human recognition and automated moderation, we organize typographic manipulations into three complementary design dimensions: typographic granularity, placement strategy, and stylistic transformation. Together, these dimensions define the typographic configuration space explored in this work. Figure~\ref{fig:pipeline} illustrates the three dimensions, which jointly capture the trade-off between human perceptibility and machine detectability.

\noindent $\bullet$ \textbf{Typographic Granularity (L)} determines how toxic content is decomposed before being embedded into surrounding benign text. Toxic expressions may be preserved as complete words (Word), segmented into token-level units (Token), or represented using mixed token--character decompositions (Mixed). Different granularities influence how readily humans and moderation systems can reconstruct the original content.

\noindent $\bullet$ \textbf{Placement Strategies (M)} determine how toxic fragments are spatially arranged within benign text. We consider a range of representative layouts, including vertical ($\text{M}_1-\text{M}_3$), diagonal ($\text{M}_4-\text{M}_5$), and randomized placements ($\text{M}_6$). These strategies alter the visual organization of toxic content and may enable human readers to perceive coherent patterns that are not explicitly reflected in the sequential text representation processed by moderation systems.

\noindent $\bullet$ \textbf{Stylistic Transformations (S)} modify the visual appearance of selected characters or tokens. We consider commonly used typographic cues such as bolding ($\text{Bold}$), color changes ($\text{Col}$), highlighting ($\text{Hi}$), capitalization ($\text{Cap}$), precomposed Unicode characters ($\text{Precomposed}$), and masking-based transformations ($\text{Colze}$). These modifications can increase the visual salience of toxic content for human readers while remaining weakly represented in tokenized text.

\begin{figure*}[h]
    \centering
    \includegraphics[width=0.96\linewidth]{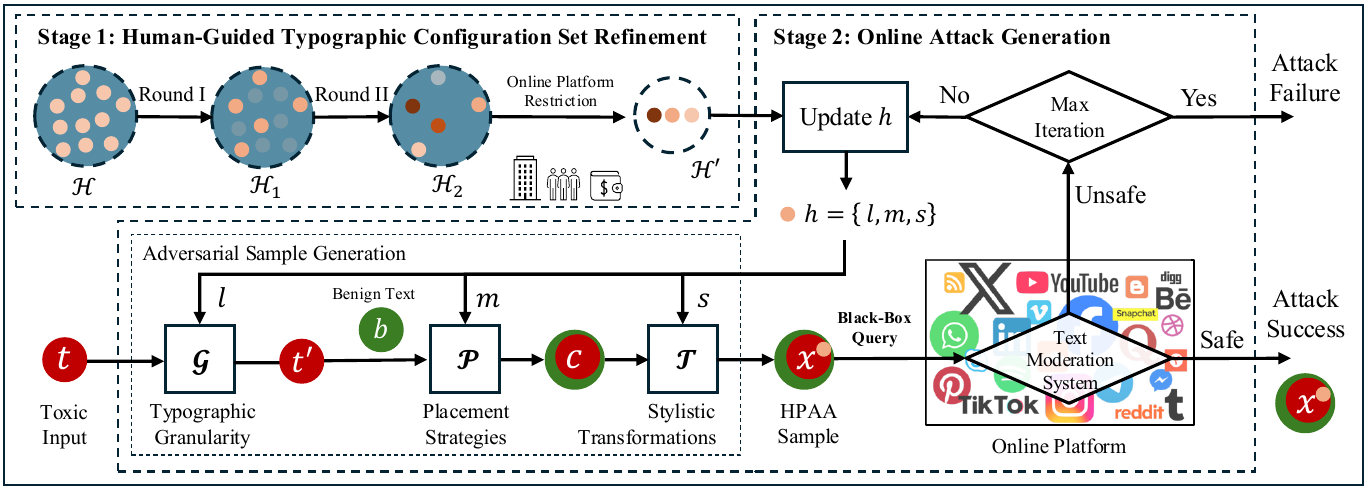}
    \caption{The HPAA framework incorporates a \emph{two-round user study} (in Stage 1 Human-Guided Refinement), conducted once to collect human feedback that informs and supports the attack design.}
    \label{fig:framework}
\end{figure*}

By combining granularity, placement, and style, HPAA generates a diverse set of typographic configurations that vary in both human perceptibility and machine detectability. Each configuration represents a distinct combination of visual cues that may differentially affect how humans and moderation systems interpret the same content. Figure~\ref{fig:pipeline} also presents an illustrative example using the M$_2$-Word-$\text{Hi}$ configuration, where toxic words are preserved at the word level, vertically arranged within benign text, and visually highlighted to remain recognizable to human readers.

\vspace{0.05in}

\noindent\textbf{Design Rationale for Stage 1.}
The configuration space defined above contains many possible cue combinations, and it is unclear a priori which configurations best preserve human recognition. Rather than relying on manual intuition, we conduct a two-round user study to systematically identify perceptually effective configurations. The resulting human-ranked configuration set is subsequently used to guide attack generation and evaluation.

\section{Procedures in HPAA}
\label{sec:framework}

Figure~\ref{fig:framework} presents the overall HPAA framework. Given a toxic input message $t$, HPAA aims to construct a typographically modified sample $x$ that remains recognizable to human readers while avoiding detection by automated moderation systems.

The key challenge is that moderation evasion and human perceptibility must be optimized simultaneously. Because the typographic configuration space contains numerous possible combinations of granularity, placement, and stylistic transformations, directly incorporating human evaluation into every attack attempt would be prohibitively expensive and difficult to scale. HPAA therefore decouples human perceptibility from moderation evasion through a two-stage framework.

In Stage~1, HPAA performs a one-time human-guided refinement process to identify typographic configurations that reliably preserve human recognition of embedded toxic content. Starting from the full typographic configuration space, we conduct a two-round user study to evaluate how different combinations of granularity, placement, and stylistic transformations affect human interpretability. The resulting configurations are subsequently filtered according to platform-specific admissibility constraints to obtain a deployable configuration set for attack generation.

In Stage~2, HPAA generates adversarial samples using configurations selected from the refined set and evaluates them through black-box interaction with the target moderation system. Given a toxic message $t$ and benign carrier text $b$, HPAA embeds the toxic content according to a selected typographic configuration and submits the resulting sample to the moderation system. Based on the returned moderation outcome, HPAA iteratively explores alternative configurations under a limited query budget until a successful moderation bypass is identified or the budget is exhausted.

By separating the expensive task of estimating human perceptibility from the online process of moderation evasion, HPAA can efficiently attack previously unseen moderation systems while preserving perceptual properties validated through human evaluation.

\subsection{Stage 1: Configuration Set Refinement}
\label{sec:stage1}

The typographic configuration space introduced in Section~\ref{sec:typographic-cues} contains numerous cue combinations with unknown perceptual effectiveness. While some configurations may appear visually salient, it is unclear whether they reliably preserve human interpretability across different users and content categories. Rather than relying on manual intuition, HPAA performs a two-round human-guided refinement process to systematically identify perceptually effective configurations and construct a refined configuration set for subsequent attack.

\textbf{Round I: Configuration Screening.}
The first round evaluates candidate configurations drawn from the full configuration space. Using pairwise comparisons, participants assess the recognizability of embedded toxic content under different configurations. This stage identifies configurations that consistently preserve human interpretability across diverse participants while mitigating individual differences in judgment, producing an initial ranking.

\textbf{Round II: Targeted Configuration Evaluation.}
Using the configurations retained from Round I, the second round conducts a more focused evaluation under controlled conditions. This stage examines recognition accuracy, partial versus complete recognition, and robustness across different content categories. Based on the resulting rankings, HPAA identifies configurations that most reliably preserve human recognizability.

To ensure practical applicability, platform-specific admissibility constraints are further applied to obtain a deployable configuration subset
$\mathcal{H}' \subseteq \mathcal{H},$
where $\mathcal{H}$ denotes the original configuration space. The resulting set $\mathcal{H}'$ contains configurations that are both perceptually effective and technically realizable on the target platform. HPAA restricts subsequent attack generation to configurations in $\mathcal{H}'$.

Detailed descriptions of the study design, participant recruitment, and analysis methodology are provided in Section~\ref{sec:user_study}. Results for both rounds are reported in Appendix~\ref{sec:user_study-results}.

\subsection{Stage 2: Online Attack Generation}
\label{sec:stage2}

Given the refined configuration set $\mathcal{H}'$, HPAA performs online attack generation through black-box interaction with the target moderation system. Unlike Stage~1, which focuses on human perceptibility, Stage~2 focuses on practical moderation evasion. By restricting the search space to configurations that have already been validated through human evaluation, HPAA can efficiently concentrate its query budget on exploring moderation weaknesses rather than repeatedly verifying perceptual effectiveness.

\subsubsection{Adversarial Sample Generation}
\label{sec:sample-generation}

Given a selected configuration $h=\{l,m,s\}\in\mathcal{H}'$, HPAA constructs an adversarial sample from toxic input $t$ and benign carrier text $b$: $x = F(t,b;h)$, where $F(\cdot)$ denotes the typographic embedding procedure that transforms toxic input $t$ into an adversarial sample according to configuration $h$ while using benign carrier text $b$.

The generation process first decomposes the toxic content according to the selected granularity level $l$. The resulting textual units are then embedded into the benign carrier text following the placement strategy $m$. Finally, the stylistic transformation $s$ is applied to introduce the corresponding visual cues. Together, these operations produce the final adversarial sample submitted to the moderation system.

Because all configurations are drawn from the human-refined set $\mathcal{H}'$, generated samples inherit the perceptual properties validated in Stage~1. Consequently, attack generation can focus on moderation evasion without repeatedly verifying human recognizability.

\subsubsection{Black-Box Moderation Interaction}
\label{sec:moderation-interaction}

For a candidate configuration $h=\{l,m,s\}\in\mathcal{H}'$, HPAA generates an adversarial sample and submits it to the target moderation system. The detector is treated as a black box, and HPAA observes only the final outcome (e.g., \textit{Safe} or \textit{Unsafe}).

The moderation outcome serves as a lightweight optimization signal for configuration selection. If the generated sample receives a benign moderation decision, the attack terminates. Otherwise, HPAA explores alternative configurations from $\mathcal{H}'$ and repeats the generation process. By leveraging the human-refined configuration set, HPAA concentrates detector queries on a substantially smaller and more promising search space. The procedure continues until a successful evasion sample is identified or the maximum query budget is reached.

\section{Attack Evaluations}
\label{sec:evaluations}

\subsection{Experimental Setup}\label{sec:experiments-evaluation-configs}

\noindent \textbf{Stylistic Transformation Grounding.} We ground the stylistic transformation dimension of HPAA in formatting conventions observed across five major real-world platforms: Reddit, X/Twitter, Discord, Stack Overflow, and YouTube. Specifically, we surveyed the native text-formatting features supported by each platform and identified transformations that can be deployed using functionality available to ordinary users without special tools or privileges. Table~\ref{tab:platforms} in Appendix summarizes the supported formatting features across platforms. Most platforms support bold and precomposed text, while some additionally support highlighting and color under varying character limits (280 -- 40k). Based on these observations, we derive six stylistic transformations as in Figure~\ref{fig:pipeline}.

\vspace{0.05in}

\noindent \textbf{Datasets.} To support the two-stage HPAA framework, we construct five datasets: a Short Toxic Text Dataset (STTD), a Benign Text Dataset (BTD), two user-study datasets (HUS-I and HUS-II), and an HPAA Evaluation Dataset (HED).

\begin{itemize}[leftmargin=*, topsep=0pt, itemsep=0em]
    \item \textbf{Short Toxic Text Dataset (STTD).} We construct a short toxic text dataset by re-annotating 249 samples from AdvBench~\cite{zou2023universal} using the Llama Guard safety taxonomy~\cite{inan2023llama}, which defines five harmful categories: Hate, Violence, Sexual Content, Self-Harm, and Insults. We retain only toxic sentences containing at most ten words.
    \item \textbf{Benign Text Dataset (BTD).} We construct a benign text corpus by sampling reviews with the maximum available rating from five public datasets: TripAdvisor (hotels)~\cite{tripadvisor2024}, Yelp (restaurants)~\cite{yelp2024}, IMDb (movies)~\cite{maas2011learning}, Amazon Product Reviews~\cite{mcauley2015image}, and Amazon Music Reviews~\cite{he2016ups}.
    \item \textbf{Dataset for HPAA User Study Round I (HUS-I).} HUS-I contains 108 samples, each formed by embedding a toxic expression into benign carrier text using a distinct typographic configuration. The dataset is used for Round~I configuration screening during Stage~1 refinement.
    \item \textbf{Dataset for HPAA User Study Round II (HUS-II).} HUS-II contains 105 samples spanning five topic domains (21 samples per domain). Each sample is generated using one configuration from the top-ranked configuration set identified in Round~I. The dataset is used for Round~II evaluation and refinement of candidate configurations.
    \item \textbf{HPAA Evaluation Dataset (HED).} HED is constructed by applying configurations from the refined configuration set $\mathcal{H}'$ to samples drawn from STTD and BTD. The dataset is used to evaluate HPAA against automated moderation systems and serves as the primary benchmark for attack effectiveness.
\end{itemize}

\vspace{0.05in}

\noindent \textbf{Evaluation Metrics.} We evaluate HPAA along two complementary dimensions: human exposure and attack effectiveness against automated moderation systems.

\begin{itemize}[leftmargin=*, topsep=0pt, itemsep=0em]
\item \textbf{Human Exposure.} We quantify human exposure using the recognition frequency measured in Round~II of the user study. This measure captures the proportion of participants who correctly identify the intended harmful content after typographic transformation and serves as a proxy for population-level exposure. Higher recognition frequencies indicate greater semantic accessibility to human observers and thus a higher potential for human harm.
\item \textbf{Attack Effectiveness on Automated Moderation.}
    \begin{itemize}[leftmargin=*, topsep=0pt, itemsep=0em]
        \item \textit{Detection and Evasion Rates.} For detectors with discrete outputs, detection rates are computed directly; for detectors producing continuous safety scores, we apply a fixed decision threshold calibrated on a balanced validation set to obtain binary predictions. We report the detectors' \emph{detection rate}, defined as the fraction of toxic samples classified as \emph{unsafe}. Attack effectiveness is quantified by the \emph{evasion rate}, defined as the complement of the detection rate on adversarial samples. Detection rates on original samples reflect baseline detector performance on toxic samples, while evasion rates on HPAA samples measure the ability to bypass automated moderation.
        \item \textit{$k$-Shot Evasion Rate.} Repeated submissions may trigger platform-level rate limits, account restrictions, or human review, making successful evasion within a small number of attempts particularly desirable in practice. To capture this constraint, we evaluate evasion under different attack budgets. For an attack budget of $k$, an attacker is allowed to sequentially evaluate up to the top-$k$ configurations from $\mathcal{H}'$ in ranked order. A sample is considered successfully evaded if any evaluated configuration is classified as \emph{safe}. We report the resulting \emph{$k$-shot evasion rate}, defined as the fraction of samples that evade detection within at most $k$ attempts. In particular, we report both \emph{1-shot evasion} ($k=1$), reflecting realistic one-attempt deployment scenarios, and \emph{3-shot evasion} ($k=3$), which measures the additional benefit obtainable from a small number of refinement attempts.
    \end{itemize}
\end{itemize}

For compact presentation, we summarize attack outcomes using the notation $\{e\%_k^{r_{\min}}\}$, where $e\%$ denotes the resulting $k$-shot evasion rate, $k$ denotes the attack budget and $r_{\min}$ denotes the minimum recognition frequency of the typographic configurations involved.

\begin{figure*}[b]
    \centering
    \includegraphics[width=\linewidth]{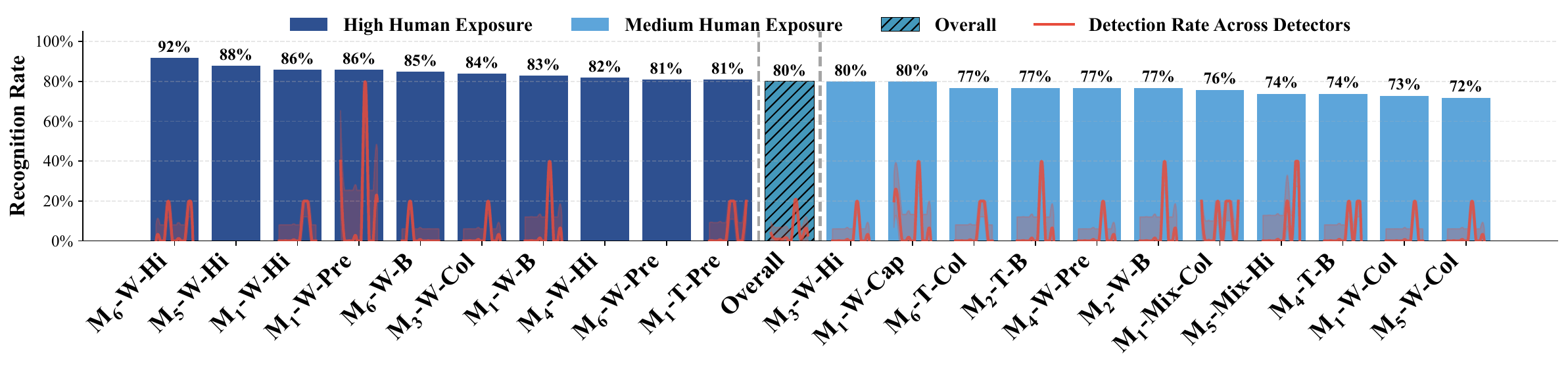}
    \caption{Human exposure rate and detectors' detection rate on HUS-II dataset. 
    }
    \label{fig:detection_rate}
\end{figure*}

\subsection{Stage 1: Configuration Set Refinement}
The typographic design space yields 108 possible configurations, collectively denoted by $\mathcal{H}$, which serve as the candidate set for Stage~1 refinement. The two refinement rounds progressively reduce this set to $\mathcal{H}_1$ and $\mathcal{H}_2$, containing 21 and 10 configurations, respectively. We discuss the resulting configuration refinement below, while the complete user-study design and procedure is presented in Section~\ref{sec:user_study-design}.

\begin{table}[H]
\centering
\scriptsize
\setlength{\tabcolsep}{4.5pt}
\renewcommand{\arraystretch}{1.05}
\caption{Human recognizability scores for the configuration set $\mathcal{H}$ on the HUS-I dataset. Based on the normalized recognition scores, the top-21 configurations are retained to form $\mathcal{H}_1$ and are shown in bold. Cells with recognizability scores $\leq 0.200$ or $\geq 0.700$ are highlighted for visual reference.}
\label{tab:round1}

\begin{tabular}{l l c c c c c c}
\toprule
\textbf{M} & \multicolumn{1}{c}{\textbf{L}}
& \textbf{Cap.} 
& \textbf{Cloze} 
& \textbf{Pre.} 
& \textbf{B.} 
& \textbf{Col.} 
& \textbf{Hi.} \\
\midrule

\multirow{3}{*}{$\text{M}_1$}
& \textbf{Word}  & \colorbox{softgreen}{\textbf{0.846}} & \colorbox{softgrey}{0.200} & \colorbox{softgreen}{\textbf{0.800}} & \colorbox{softgreen}{\textbf{0.929}} & \colorbox{softgreen}{\textbf{0.900}} & \colorbox{softgreen}{\textbf{0.800}} \\
& \textbf{Token} & 0.400 & \colorbox{softgrey}{0.067} & \colorbox{softgreen}{\textbf{0.800}} & \colorbox{softblue}{0.700} & 0.429 & \colorbox{softblue}{0.700} \\
& \textbf{Mixed} & 0.500 & \colorbox{softgrey}{0.200} & \colorbox{softblue}{0.750} & 0.583 & \colorbox{softgreen}{\textbf{0.909}} & \colorbox{softblue}{0.727} \\
\midrule

\multirow{3}{*}{$\text{M}_2$}
& \textbf{Word}  & 0.538 & \colorbox{softgrey}{0.154} & 0.375 & \colorbox{softgreen}{\textbf{1.000}} & 0.692 & \colorbox{softblue}{0.750} \\
& \textbf{Token} & 0.375 & \colorbox{softgrey}{0.167} & 0.667 & \colorbox{softgreen}{\textbf{0.909}} & 0.462 & \colorbox{softblue}{0.750} \\
& \textbf{Mixed} & 0.353 & \colorbox{softgrey}{0.143} & \colorbox{softgrey}{0.200} & 0.286 & 0.667 & 0.500 \\
\midrule

\multirow{3}{*}{$\text{M}_3$}
& \textbf{Word}  & 0.364 & 0.333 & 0.636 & \colorbox{softblue}{0.727} & \colorbox{softgreen}{\textbf{0.909}} & \colorbox{softgreen}{\textbf{0.933}} \\
& \textbf{Token} & \colorbox{softgrey}{0.200} & 0.333 & \colorbox{softblue}{0.700} & 0.583 & \colorbox{softblue}{0.778} & 0.500 \\
& \textbf{Mixed} & \colorbox{softgrey}{0.167} & 0.222 & \colorbox{softblue}{0.700} & 0.591 & 0.692 & 0.636 \\
\midrule

\multirow{3}{*}{$\text{M}_4$}
& \textbf{Word}  & 0.500 & \colorbox{softgrey}{0.000} & \colorbox{softgreen}{\textbf{0.909}} & \colorbox{softblue}{0.727} & 0.636 & \colorbox{softgreen}{\textbf{0.909}} \\
& \textbf{Token} & 0.300 & \colorbox{softgrey}{0.071} & 0.600 & \colorbox{softgreen}{\textbf{0.800}} & 0.615 & \colorbox{softblue}{0.750} \\
& \textbf{Mixed} & 0.333 & \colorbox{softgrey}{0.182} & 0.417 & 0.667 & 0.417 & 0.667 \\
\midrule

\multirow{3}{*}{$\text{M}_5$}
& \textbf{Word}  & 0.538 & 0.250 & 0.733 & 0.636 & \colorbox{softgreen}{\textbf{0.875}} & \colorbox{softgreen}{\textbf{0.875}} \\
& \textbf{Token} & 0.455 & \colorbox{softgrey}{0.133} & 0.636 & 0.692 & 0.429 & 0.429 \\
& \textbf{Mixed} & 0.375 & \colorbox{softgrey}{0.133} & 0.455 & 0.667 & 0.300 & \colorbox{softgreen}{\textbf{0.818}} \\
\midrule

\multirow{3}{*}{$\text{M}_6$}
& \textbf{Word}  & 0.556 & 0.400 & \colorbox{softgreen}{\textbf{0.900}} & \colorbox{softgreen}{\textbf{0.800}} & 0.500 & \colorbox{softgreen}{\textbf{0.824}} \\
& \textbf{Token} & 0.471 & 0.313 & 0.583 & \colorbox{softblue}{0.714} & \colorbox{softgreen}{\textbf{0.846}} & 0.636 \\
& \textbf{Mixed} & 0.667 & 0.600 & 0.545 & 0.545 & 0.545 & 0.364 \\

\bottomrule
\end{tabular}
\end{table}

\newcommand{\rar}{\(\textcolor{red}{\downarrow}\)}

\begin{table*}[htbp]
\centering
\footnotesize
\caption{
Comparison of content safety model detection rates before and after applying HPAA across five harmful categories. Results are reported on the STTD dataset (Original) and the HUS-II dataset (HPAA-applied) with iteration number $k=1$ and $\mathcal{H}'=\mathcal{H}_2$ (human exposure rate: 92\%). Detection rates lower than \colorbox{softgreen}{1\%} and higher than \colorbox{softred}{10\%} are highlighted.
}
\label{tab:model_robustness}
\renewcommand{\arraystretch}{1}
\setlength{\tabcolsep}{3pt}

\begin{tabular}{l|ccc|ccc|ccc|ccc|ccc}
\hline
\textbf{Model Name} & \multicolumn{3}{c|}{\textbf{Hate}} & 
\multicolumn{3}{c|}{\textbf{Violent}} & 
\multicolumn{3}{c|}{\textbf{Sexual}} & 
\multicolumn{3}{c|}{\textbf{Self-harm}} & 
\multicolumn{3}{c}{\textbf{Insults}} \\ 
\cline{2-16}
 & Original & HPAA &  & Original & HPAA &  & Original & HPAA &  & Original & HPAA &  & Original & HPAA &  \\ 
\hline

SG-9B
& 54.41\% & \colorbox{softgreen}{0.00\%} & \rar 
& 53.51\% & \colorbox{softgreen}{0.00\%} & \rar 
& 66.67\% & \colorbox{softgreen}{0.00\%} & \rar 
& 75.00\% & \colorbox{softgreen}{0.00\%}& \rar 
& 44.93\% & \colorbox{softgreen}{0.00\%} & \rar \\

PA
& 77.94\% & 1.52\% & \rar 
& 71.05\% & \colorbox{softgreen}{0.89\%} & \rar 
& 94.44\% & 5.88\% & \rar 
& 79.55\% & 2.33\% & \rar 
& 61.59\% & 1.52\% & \rar \\

Amazon-C
& 94.12\% & 1.47\% & \rar 
& 94.74\% & \colorbox{softgreen}{0.00\%} & \rar 
& 97.22\% & 2.78\% & \rar 
& 93.18\% & 4.55\% & \rar 
& 78.26\% & 3.62\% & \rar \\

Enkrypt AI 
& 80.88\% & \colorbox{softgreen}{0.00\%} & \rar 
& 72.81\% & \colorbox{softgreen}{0.00\%} & \rar 
& 97.22\% & \colorbox{softgreen}{0.00\%} & \rar 
& 84.09\% & \colorbox{softgreen}{0.00\%} & \rar 
& 65.22\% & \colorbox{softgreen}{0.00\%} & \rar \\

Azure AI 
& 86.76\% & \colorbox{softred}{11.76\%} & \rar 
& 82.46\% & \colorbox{softred}{14.04\%} & \rar 
& 86.11\% & \colorbox{softred}{16.67\%} & \rar 
& 88.64\% & \colorbox{softred}{13.64\%} & \rar 
& 60.87\% & \colorbox{softred}{11.59\%} & \rar \\

LG3-8B
& 91.18\% & \colorbox{softgreen}{0.00\%} & \rar 
& 89.47\% & \colorbox{softgreen}{0.00\%} & \rar 
& 88.89\% & \colorbox{softgreen}{0.00\%} & \rar 
& 97.73\% & \colorbox{softgreen}{0.00\%} & \rar 
& 74.64\% & \colorbox{softgreen}{0.00\%} & \rar \\

G2F
& 100.00\% & 1.47\% & \rar 
& 89.47\% & \colorbox{softgreen}{0.00\%} & \rar 
& 91.67\% & \colorbox{softgreen}{0.00\%} & \rar 
& 100.00\% & \colorbox{softgreen}{0.00\%} & \rar 
& 75.36\% & \colorbox{softgreen}{0.72\%} & \rar \\

GPT-4o 
& 98.53\% & 4.41\% & \rar 
& 94.74\% & 2.63\% & \rar 
& 97.22\% & 2.78\% & \rar 
& 100.00\% & 2.27\% & \rar 
& 89.13\% & 2.90\% & \rar \\

G-2.5-FL
& 100.00\% & \colorbox{softred}{11.76\%} & \rar 
& 98.25\% &  7.89\% & \rar 
& 97.22\% & 8.33\% & \rar 
& 100.00\% & 6.82\% & \rar 
& 92.03\% & \colorbox{softred}{10.14\%} & \rar \\

Omni
& 95.59\% & 2.94\% & \rar 
& 92.98\% & 1.75\% & \rar 
& 97.22\% & 2.78\% & \rar 
& 90.91\% & \colorbox{softgreen}{0.00\%} & \rar 
& 77.54\% & 1.45\% & \rar \\

\hline
\end{tabular}
\normalsize
\end{table*}

\noindent \textbf{$\mathcal{H}_1$ from Round I.} Table~\ref{tab:round1} reports the normalized recognition scores (scaled to $[0,1]$) derived from the Round~I pairwise-comparison study, from which we retain the top-21 configurations to form $\mathcal{H}_1$. Since the benign and toxic content were fixed throughout Round~I and only the configuration varied, the pairwise-comparison refinement isolates the effect of typographic design on recognition performance. This controlled setup provides an efficient means of estimating relative preferences among a large number of configurations, thereby enabling effective exploration of the configuration space $\mathcal{H}$.

Specifically, configurations employing visually salient transformations such as \textit{Bold}, \textit{Color}, \textit{Highlight}, and \textit{Precomposed} consistently achieve higher recognition scores than \textit{Cloze}-based variants, which intentionally obscure lexical content and therefore reduce recognizability. We further observe that placement strategy $\text{M}_1$ and word-level granularity generally yield stronger recognition performance, suggesting that preserving lexical coherence and spatial regularity facilitates human reconstruction of toxic content.

Overall, the results reveal a clear trade-off between concealment and recognizability. These findings provide a principled basis for pruning $\mathcal{H}$ and constructing the refined set $\mathcal{H}_1$.

\vspace{0.05in}

\noindent \textbf{$\mathcal{H}_2$ from Round II.} While the normalized scores obtained in Round~I capture relative differences among configurations, they do not directly quantify how often toxic content is recognized by human readers. We therefore conduct Round~II to measure recognition frequency, defined as the fraction of participants who correctly identify the embedded toxic content under each configuration. This metric serves as a proxy for human perceptibility and population-level exposure. Details of the computation are provided in Appendix~\ref{sec:user_study-results}.

To keep the human evaluation tractable, Round~II evaluates the configurations retained in $\mathcal{H}_1$. As shown in Figure~\ref{fig:detection_rate}, several configurations achieve recognition frequencies above 80\% while simultaneously exhibiting low detection rates across detectors, indicating a persistent mismatch between human perception and automated moderation. Configurations ranked highly in Round~I generally remain highly recognizable in Round~II, suggesting that the pairwise-comparison step provides a stable approximation of perceptual accessibility.

The results further show that human recognizability and detector sensitivity are only weakly aligned. Content that is readily understood by human participants frequently evades moderation systems, whereas configurations receiving comparatively stronger detector responses often remain highly recognizable.

Accordingly, we retain configurations achieving at least 80\% recognition frequency and construct the final refined set $\mathcal{H}_2$, containing 10 highly recognizable configurations for subsequent attack generation.

\subsection{Stage 2: Online Attack Generation}

\vspace{-0.02in}
\textbf{1-Shot Evasion.} Table~\ref{tab:model_robustness} reports detector performance under a 1-shot attack setting, where HPAA uses the highest-ranked configuration in $\mathcal{H}'$ (M$_6$-Word-Hi), with a 92\% human recognition rate. This setting reflects a realistic deployment scenario in which an attacker attempts to evade moderation using a single submission. HPAA consistently reduces detection rates across all evaluated moderation systems and harmful-content categories, often driving detection rates to near zero. Although Azure Content Safety and Gemini~2.5 Flash Lite are comparatively more robust, their post-attack detection rates still remain below 17\% across all categories. HPAA reduces detection rates by one to two orders of magnitude while maintaining high human recognizability. Across all categories and moderation systems, the strongest HPAA configuration achieves up to $\{83.33\%_1^{92\%}\}$, corresponding to an evasion rate of 83.33\% with a 92\% human recognition rate. \footnote{Detector abbreviations: LG3-8B (Llama-Guard-3-8B~\cite{chi2024llama}), G2F (Gemini 2.0 Flash~\cite{comanici2025gemini}), G-2.5-FL (Gemini 2.5 Flash-Lite~\cite{comanici2025gemini}), SG-2B/SG-9B (ShieldGemma 2B/9B~\cite{google2025safermultimodal}), Azure (Azure AI Content Safety API~\cite{microsoft2025azurecontentsafety}), PA (Perspective API~\cite{muralikumar2023human}), Amazon-C (Amazon Comprehend Toxicity Detection~\cite{aws2023newcomprehendtoxicity}), Amazon-N (Amazon Nova Lite 2.0~\cite{aws2023detecttoxiccontent}), Omni (Omni-Moderation-Latest~\cite{li2025baichuan_omni}), Enkrypt AI (Enkrypt AI Guardrails API~\cite{enkrypt2025unifiedguardrails}), GPT-3.5 (ChatGPT-3.5~\cite{openai2025o3o4mini}), and GPT-4o (GPT-4o~\cite{openai2024gpt4omini}).}

\vspace{0.02in}

\begin{figure*}[htbp]
    \centering
    \begin{subfigure}[b]{0.48\textwidth}
        \centering
        \includegraphics[width=\textwidth]{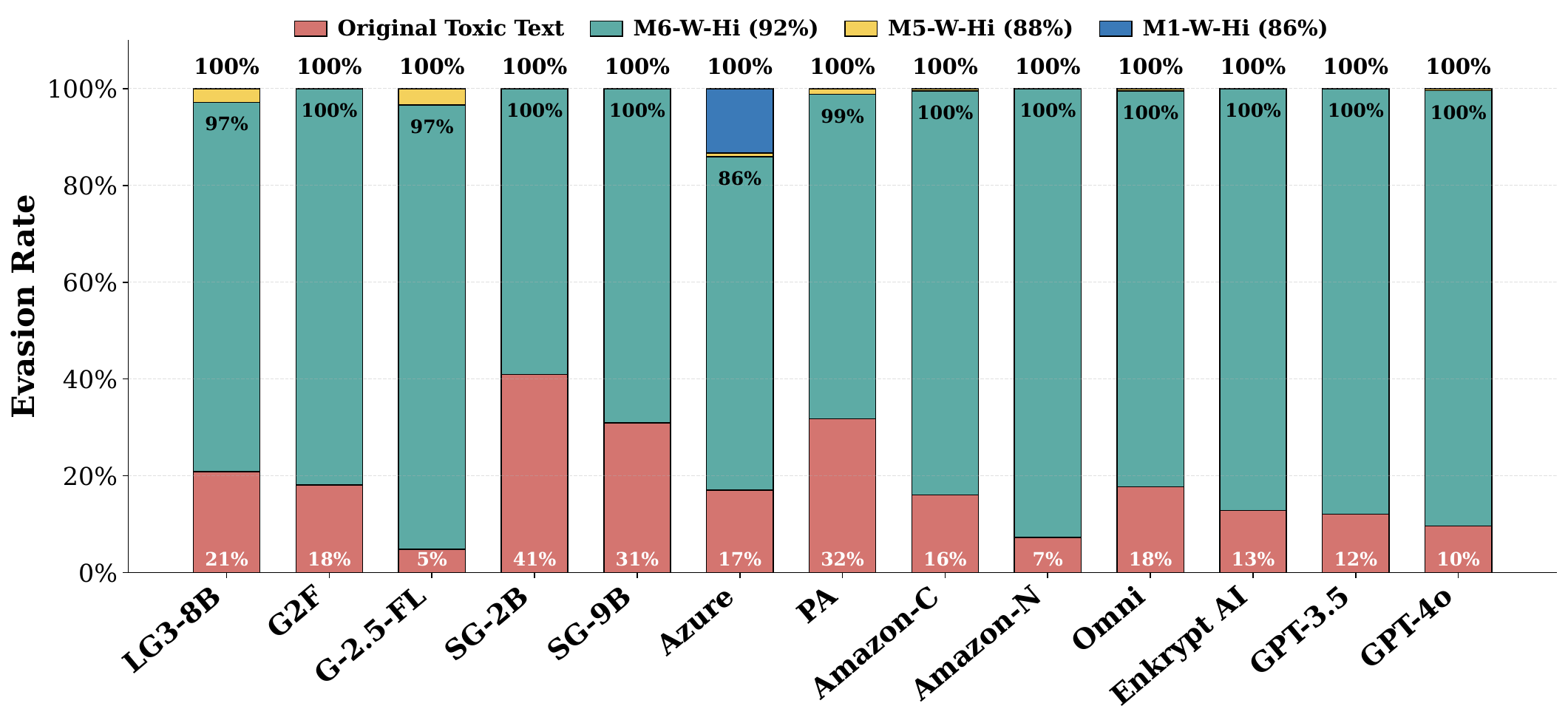}
        \caption{The evasion rate on STTD and HED dataset when $\mathcal{H}'=\mathcal{H}_2$, i.e., when the highlight is allowed (e.g., Stack Overflow).
        }
        \label{fig:dr_53}
    \end{subfigure}
    \hfill
    \begin{subfigure}[b]{0.48\textwidth}
        \centering
        \includegraphics[width=\textwidth]{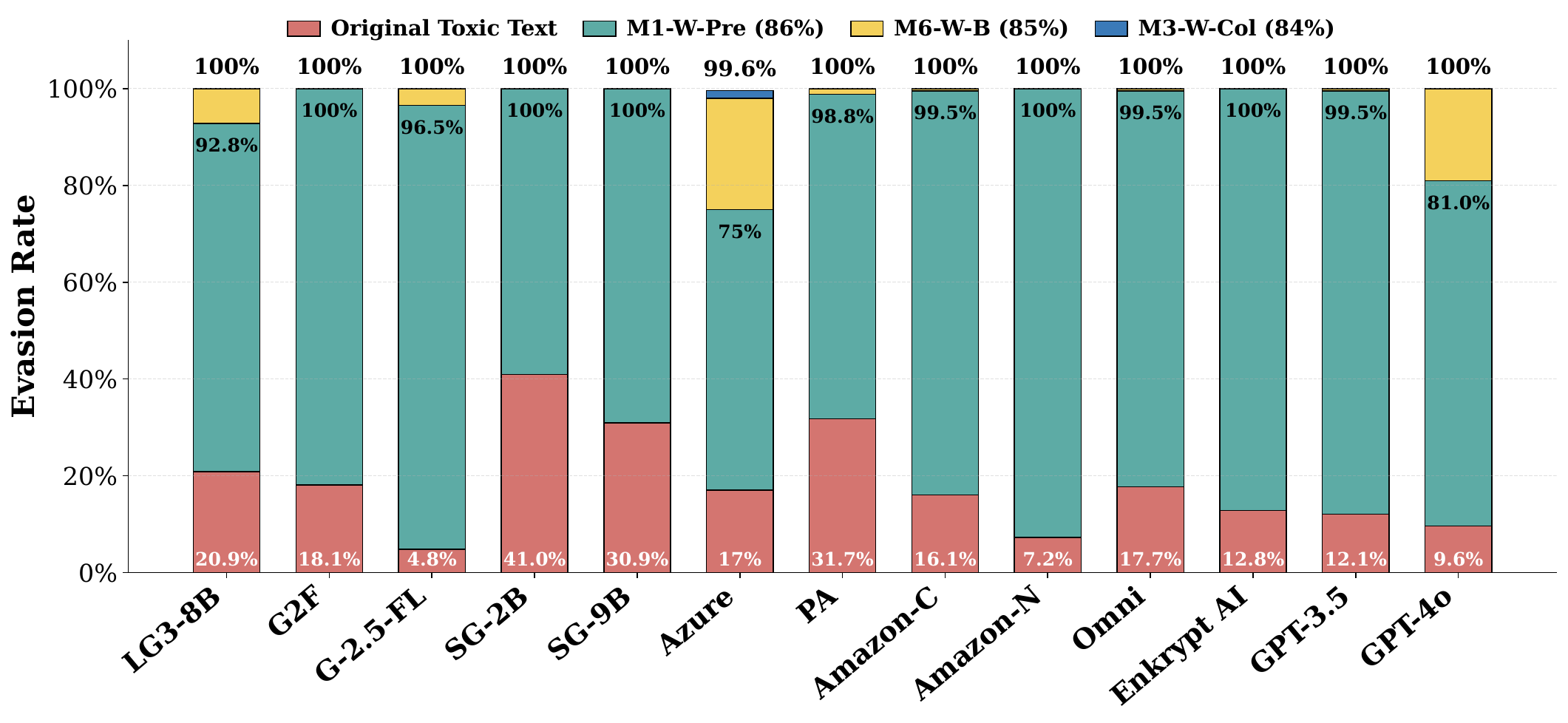}
        \caption{The evasion rate on STTD and HED dataset when $\mathcal{H}'\subset\mathcal{H}_2$, i.e., when the highlight is not allowed (e.g., Reddit).}
        \label{fig:dr_54}
    \end{subfigure}
    \caption{The evasion results on STTD (red shadow) and HED datasets (3 iterations).}
    \label{fig:evasion_results}
\end{figure*}

\noindent \textbf{3-Shot Evasion.} Figure~\ref{fig:evasion_results} reports detector performance under a larger attack budget ($k=3$), where an adversary may sequentially evaluate up to three configurations from $\mathcal{H}'$. We evaluate detector robustness under both the full configuration set ($\mathcal{H}'=\mathcal{H}_2$) and a restricted setting ($\mathcal{H}'\subset\mathcal{H}_2$) where highlighting is unavailable.

Across all evaluated systems, later attack attempts continue to provide substantial gains despite operating only on samples that survive earlier attacks, indicating that different configurations exploit complementary moderation weaknesses. With a 3-shot budget, HPAA achieves up to $\{100\%_3^{86\%}\}$ (or $\{100\%_3^{84\%}\}$ under the restricted setting), demonstrating that near-complete moderation evasion can be achieved with only a small number of attempts.

\subsection{Sensitivity Analysis}

\subsubsection{Effect of Decision Thresholds}

\begin{figure}[H]
    \centering
    \includegraphics[width=\linewidth]{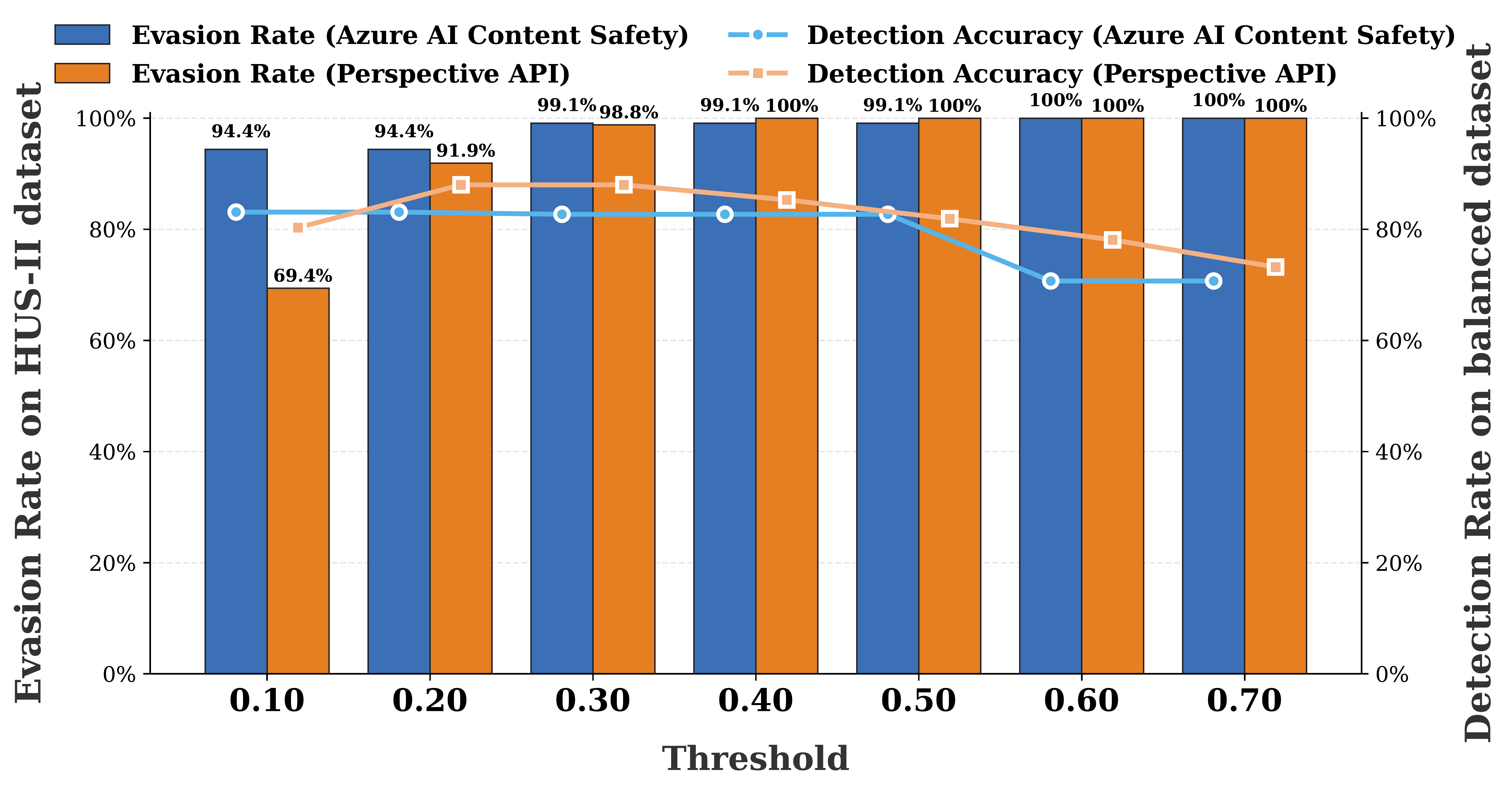}
    \caption{Sensitivity of threshold-based moderation systems.
    }
    \label{fig:threshold}
\end{figure}

For threshold-based moderation systems that rely on continuous safety scores (e.g., \textit{Azure Content Safety} and \textit{Perspective API}), we use a default decision threshold of 0.5, which serves as a commonly adopted baseline when no task-specific threshold is prescribed. Our sensitivity analysis shows that while alternative thresholds can yield marginal performance variations, even optimally tuned thresholds provide limited effectiveness against our attack setting. As illustrated in Figure~\ref{fig:threshold}, we report the detection accuracy to reflect the classification capability of the detectors on a balanced evaluation set, which is constructed by uniformly sampling an equal number of benign and toxic sentences from the HED dataset. The evasion rate is measured on the HPAA samples in HED dataset and quantifies the effectiveness of HPAA in bypassing automated moderation under varying decision thresholds. We report the detection accuracy to show the classification ability of detectors lowering the decision threshold increases evasion rates but this comes at the cost of substantially reduced detection accuracy on the balanced dataset.
As a result, aggressive threshold tuning is impractical in deployment and offers only limited robustness against HPAA-based evasion.

\subsubsection{Effect of Benign Content Ratio}

\begin{figure}[H]
    \centering
    \begin{subfigure}{\linewidth}
        \centering
        \includegraphics[width=0.8\linewidth]{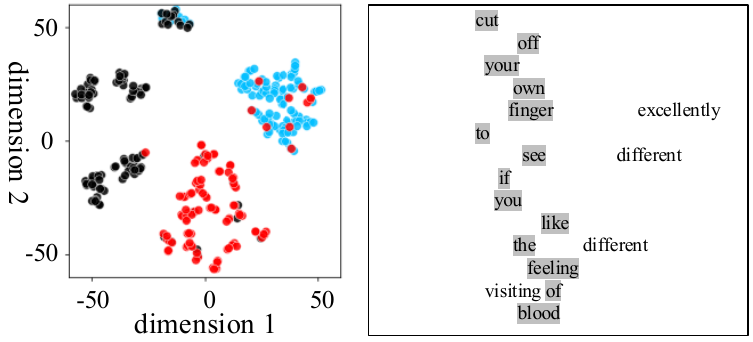}
        \caption{10\% benign words retained; overlap with toxic clusters.}
        \label{fig:demo.1}
    \end{subfigure}
    \begin{subfigure}{\linewidth}
        \centering
        \includegraphics[width=0.8\linewidth]{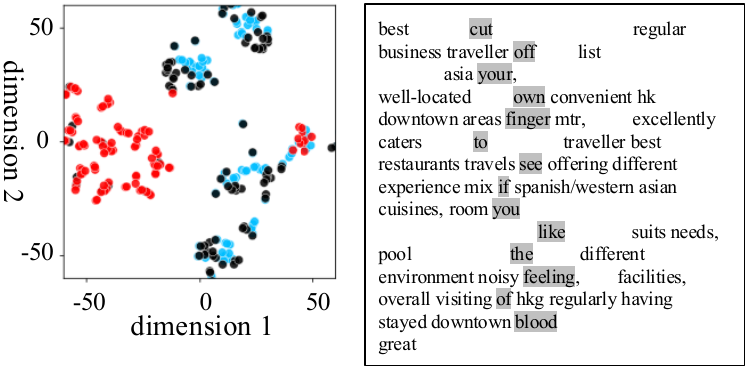}
        \caption{70\% benign words retained; intermediate semantic alignment.}
        \label{fig:demo.2}
    \end{subfigure}
    \begin{subfigure}{\linewidth}
        \centering
        \includegraphics[width=0.8\linewidth]{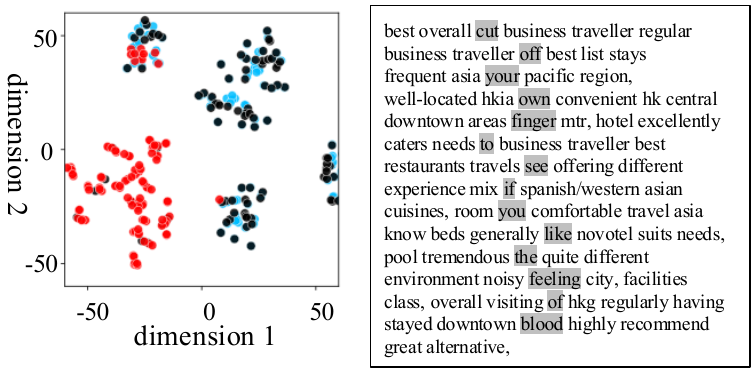}
        \caption{90\% benign words retained; alignment with benign clusters.}
        \label{fig:demo.3}
    \end{subfigure}
    \caption{Effect of benign word ratio on HPAA samples. \\
Left: t-SNE embeddings (\textcolor{red}{Red}: Toxic Text Only, \textcolor{blue}{Blue}: Benign Text Only, \textbf{Black}: HPAA samples); right: text examples.}
    \label{fig:demo}
\end{figure}

We analyze HPAA's sensitivity to the amount of benign content inserted into toxic text. In Figure~\ref{fig:demo}, reducing the benign content causes HPAA samples to move closer to toxic regions in embedding space, while increasing benign content shifts them toward benign clusters.

\subsubsection{Effect of Stylistic Transformation}

Stylistic transformations play a critical role in HPAA effectiveness. Simply interleaving toxic content into benign text already reduces detector performance, achieving approximately 70\% evasion on the HED dataset against Perspective API. However, incorporating typographic cues further amplifies the perceptual mismatch exploited by HPAA. For instance, the M$_6$-Word-Hi configuration increases evasion to 99\%, indicating that visually salient transformations substantially improve moderation evasion while preserving human recognizability. This result suggests that attack effectiveness is not solely attributable to the presence of benign carrier text, but also depends on the visual cues used to guide human interpretation.

\begin{table*}[htbp]
\centering
\caption{Evasion rates on Advbench with M$_\text{6}$-W-Hi evaluated by Perspective API.}
\label{tab:perspective}
\resizebox{\textwidth}{!}{
\begin{tabular}{@{}lcccccccccccc@{}}
\toprule
Toxic Length & Avg & 0.01 & 0.05 & 0.1 & 0.2 & 0.3 & 0.4 & 0.5 & 0.6 & 0.7 & 0.8 & 0.9 \\
\midrule
$\leq$5 words  & 98.0\% & 97.3\% & 97.3\% & 97.2\% & 97.3\% & 97.3\% & 97.2\% & 97.3\% & 100.0\% & 97.2\% & 100.0\% & 100.0\% \\
6--10 words    & 96.5\% & 97.3\% & 94.6\% & 100.0\% & 97.2\% & 97.3\% & 97.2\% & 94.6\% & 94.6\% & 97.3\% & 94.6\% & 97.3\% \\
11--15 words   & 100.0\% & 100.0\% & 100.0\% & 100.0\% & 100.0\% & 100.0\% & 100.0\% & 100.0\% & 100.0\% & 100.0\% & 100.0\% & 100.0\% \\
16--20 words   & 100.0\% & 100.0\% & 100.0\% & 100.0\% & 100.0\% & 100.0\% & 100.0\% & 100.0\% & 100.0\% & 100.0\% & 100.0\% & 100.0\% \\
\bottomrule
\end{tabular}%
}
\end{table*}

\begin{table*}[htbp]
\centering
\caption{Evasion rates on Advbench with M$_\text{6}$-W-Hi evaluated by Azure Content Moderator.}
\label{tab:azure}
\resizebox{\textwidth}{!}{%
\begin{tabular}{@{}lcccccccccccc@{}}
\toprule
Toxic Length  & Avg & 0.01 & 0.05 & 0.1 & 0.2 & 0.3 & 0.4 & 0.5 & 0.6 & 0.7 & 0.8 & 0.9 \\
\midrule
$\leq$5 words                     & 98.8\% & 97.3\%  & 100.0\% & 97.3\%  & 100.0\% & 97.3\%  & 100.0\% & 94.6\%  & 100.0\% & 100.0\% & 100.0\% & 100.0\% \\
6--10 words    & 98.8\% & 94.6\%  & 100.0\% & 94.6\%  & 100.0\% & 100.0\% & 100.0\% & 100.0\% & 100.0\% & 100.0\% & 97.3\%  & 100.0\% \\ 
11--15 words & 99.3\% & 97.3\%  & 94.6\%  & 100.0\% & 100.0\% & 100.0\% & 100.0\% & 100.0\% & 100.0\% & 100.0\% & 100.0\% & 100.0\% \\ 
16--20 words & 98.8\% & 91.9\%  & 97.3\%  & 100.0\% & 100.0\% & 97.3\%  & 100.0\% & 100.0\% & 100.0\% & 100.0\% & 100.0\% & 100.0\% \\ 
\bottomrule
\end{tabular}%
}
\end{table*}

\begin{table*}[b]
\centering
\caption{Evasion rates of different attack methods across detectors.}
\label{tab:Comparison}
\resizebox{\textwidth}{!}{%
\begin{tabular}{@{}lcccccccccccccc@{}}
\toprule
Method & LG3-8B & G2F & G-2.5-FL & SG-2B & SG-9B & Azure & PA & Amazon-C & Amazon-N & Omni & Enkrypt & GPT-3.5 & GPT-4o \\
\midrule
DeepWordBug             & 22.5\% & 20.1\% & 5.2\% & 62.3\% & 38.2\% & 32.9\% & 32.4\% & 17.3\% & 9.6\%  & 18.5\% & 25.7\% & 10.4\% & 9.2\%  \\
TextBugger              & 21.9\% & 18.9\% & 5.2\% & 62.3\% & 37.8\% & 32.9\% & 32.5\% & 16.9\% & 8.4\%  & 18.9\% & 24.9\% & 10.4\% & 7.6\%  \\
TextFooler (Homoglyph)  & 23.3\% & 21.7\% & 7.6\% & 62.6\% & 38.2\% & 33.7\% & 32.8\% & 18.1\% & 12.4\% & 19.3\% & 26.9\% & 13.2\% & 10.0\% \\
\bottomrule
\end{tabular}%
}
\end{table*}

\vspace{-0.05in}

\subsection{Scaling Ability}

HPAA generalizes beyond short toxic expressions. Tables~\ref{tab:perspective}--\ref{tab:azure} and Table~\ref{tab:evasion_length} report evasion rates across toxic sentence lengths ranging from fewer than five words to twenty words and under varying benign-to-toxic ratios. Across both Perspective API and Azure Content Safety, evasion rates remain consistently high regardless of sentence length or embedding ratio, indicating that HPAA is robust to variations in content length and benign-context density.

\begin{table}[H]
\centering
\caption{Evasion rates vs.\ toxic sentence length.}
\label{tab:evasion_length}
\footnotesize
\begin{tabular}{@{}lcccc@{}}
\toprule
Toxic Words & $\leq$5 & 6--10 & 11--15 & 16--20 \\
\midrule
Perspective API & 100\%  & 99.1\% & 100\% & 100\% \\
Azure           & 97.3\% & 95.8\% & 98.1\% & 100\% \\
\bottomrule
\end{tabular}
\end{table}

\vspace{-0.2in}

\subsection{Comparison Results}

Table~\ref{tab:Comparison} compares HPAA against representative text-based adversarial attacks, including DeepWordBug, TextBugger, and TextFooler. For a fair comparison, we implement these attacks using the TextAttack framework and evaluate them on the same toxic content and moderation systems used in our experiments.

Conventional adversarial text attacks achieve only moderate success against modern moderation systems, whereas HPAA consistently attains substantially higher evasion rates while maintaining high human recognizability. Unlike existing attacks that rely on lexical perturbations, HPAA exploits a perceptual mismatch between human interpretation and automated moderation. As a result, even detectors that are relatively robust to traditional adversarial attacks remain vulnerable to human-perceptible typographic manipulations.
\section{How Do Human Perceive Toxic Phrases?}
\label{sec:user_study}

\PP{Motivation}
To address our first research question (RQ1: \textit{To what extent can human readers recognize toxic content embedded within typographically manipulated text, and how does recognition vary across different typographic configurations?}), our user study examines how human perceive and respond to adversarially perturbed textual content within social media contexts. 
While prior research has predominantly emphasized algorithmic robustness against adversarial language attacks~\cite{wallace2019universal, morris2020textattack}, the human perceptual dimension of these manipulations remains underexplored~\cite{dyrmishi2023humans}. Our study addresses this critical gap by evaluating users' ability to recognize toxic or aggressive intent in text that has been manipulated to evade large language model (LLM)-based moderation systems.

We investigate three primary dimensions: (1) the extent to which non-expert users (\textit{i.e.}, typical users) can detect harmful intent embedded in adversarial text that preserves readability, (2) the influence of typographic cue strategies (such as capitalization flipping, spatial disruption, and typographic variation) on users' perception and interpretation, and (3) how perturbation length affects human recognition patterns. 
By systematically varying both perturbation strategies and their extent, we provide a comprehensive assessment of human judgment capabilities when confronted with content designed to appear innocuous to automated systems.

Through direct comparison of user performance with LLM-based moderation systems, this investigation captures both cognitive and behavioral aspects of human interaction with adversarial content. As Morris et al.~\cite{morris2020reevaluating} emphasize, neglecting human-perceived variability in adversarial examples risks underestimating their real-world impact. Our findings can reveal potential vulnerabilities in human judgment that adversaries may exploit in real-world moderation environments and provide empirical evidence for developing more resilient content moderation mechanisms.

\subsection{User Study Design and Procedure} 
\label{sec:user_study-design}

The study was conducted in two rounds (Round I and Round II), each with distinct participant samples and experimental constraints. Both rounds were implemented as online surveys and are publicly available at \href{https://osf.io/tn2vw/overview?view_only=af5cc0f70492497ca773b58155a333c2}{User Study Access}. 

The study employed two datasets, HUS-I and HUS-II, constructed from the
Short Toxic Text Dataset (STTD) and the Benign Text Dataset (BTD)
described in Section~\ref{sec:experiments-evaluation-configs}. 
This construction ensures that differences in participant responses can be attributed primarily to the typographic manipulation rather than semantic variations in the surrounding text.

The first round examined participant' perceptual and behavioral responses to toxic language modified with 108 typographic cues. 
Analysis of participant responses identified 21 configurations that consistently preserved human recognition of toxic content. These configurations formed the hypothesis set $H_1$ and were selected for further investigation.
Building on such results, the second round systematically evaluated configurations derived from $H_1$ under controlled conditions across multiple content domains. The goal of Round II was to better understand the contextual and linguistic factors influencing toxicity recognition and to identify the top-performing configurations that formed the refined hypothesis set $H_2$. Both studies were approved by the Institutional Review Board (IRB).

\subsubsection{First Round of User Study}
\label{sec:user_study-design_6_1_1}

The first round used a structured online survey with two versions to examine which typographic cue sets aided detection of concealed harmful content across devices (smartphones and computers). 
Both versions shared identical structure, and participants in the computer-based survey were excluded from the smartphone version to prevent learning effects. 
Each survey included four sections with 30 questions in total and was designed to be completed within 10 to 15 minutes.

\vspace{0.05in}

\noindent\textbf{Embedded Text Sample Selection}. We selected toxic comments from a publicly available dataset -- AdvBench\cite{zou2023universal}. 
As these sources differ in terms of demographics and moderation standards, they provide a broader perspective on how toxicity manifests itself across platforms. For AdvBench, we extracted toxic-labeled English comments that ranged from 10 to 30 words in length, excluding duplicate or ambiguous entries. We mapped all comments to a unified binary label (``toxic'') and incorporated them into a simulated social media interface resembling a comment thread. To systematically conceal toxic phrases, we applied 108 distinct typographic cues which from a full factorial combination of three dimensions. 6 modes, 3 linguistic granularities (word, token, mixed) for each mode, and 6 visual alteration styles (cap-flipping, cloze, precomposed, bold, color, and highlight).

\vspace{0.05in}

\noindent\textbf{Structure of the Survey}. The first-round user study survey comprised three main sections, followed by a final section for completion-code verification. 
The first section presented an informed-consent form describing the study's purpose, potential risks, and data-handling procedures. 
The second section collected 9 demographic variables, including participants' country of origin, age, gender, native language, education level and frequency of social-media use. Non-native English speakers were additionally asked, “How long have you been learning English?”; this question was automatically omitted for native English speakers or who identified English as their first language.

The third section (\textit{i.e.}, main section) contains 20 questions designed to ask participants to identify which method of concealing toxic or aggressive content (made by our typographic cues) is easier to recognize. 
Each question displayed two short text passages (labeled ``A'' and ``B''), which represents different typographic cues applied to the same underlying harmful content.
Participants were asked to indicate which passage made the toxic content easier to recognize. In addition to choosing either ``A'' or ``B'', three neutral alternatives were provided: (1) ``Both texts are equally easy to recognize as toxic'', (2) ``Neither text appears to contain toxic or aggressive content'', and (3) ``I am unable to determine based on the content provided.''
Each question thus offered five response options in total. An invisible timer recorded participants' response latency for subsequent behavioral analysis. An example question of Round I from our survey was provided in Figure~\ref{fig:question} in the Appendix~\ref{sec:appendix-examples}.

The objective of Round I is to compare the relative effectiveness of typographic transformations in preserving human recognition of toxic content. In each trial, participants are presented with two passages containing the same toxic content but rendered using different typographic transformations and are asked to determine which presentation makes the harmful content easier to recognize. Because this objective is inherently comparative, we adopt a pairwise-comparison design \cite{bradley1952rank} rather than an absolute rating scheme. This allows relative judgments to be aggregated into latent recognizability rankings while minimizing inter-participant scale bias. Across all typographic cues, we generated 2,160 combined paragraphs (each composed of one passage and one harmful sentence). To maintain sampling balance, every combined passage was randomly presented at least five times across participants. The randomized pairing ensured that no rule consistently appeared in a fixed order or with a specific counterpart.


To ensure variation in textual coherence and mitigate potential bias, we mixed random and non-random passages across the survey. Non-random passages consisted of coherent, meaningful sentences or short paragraphs resembling authentic reviews or social-media posts. In contrast, random passages were syntactically or semantically incoherent text segments, created by randomly shuffling or concatenating words without meaningful context. Each survey included 40 passages (20 random, 20 non-random) presented in randomized order, and all adversarial texts were syntactically complete and semantically meaningful.

\PP{Priming Effect}
We acknowledge that our study design may introduce a priming effect because participants were informed that the study involved potentially harmful language and were asked to evaluate whether text contained harmful or aggressive content. Such awareness may increase vigilance relative to everyday use. Therefore, our recognition rates should be interpreted as measuring participant performance under informed and attentive conditions.

To further address whether participant responses were driven by the study warning, task framing, or interface design rather than the presence of harmful content, we conducted two benign-only baseline surveys. In both surveys, all samples contained benign text only, while preserving the same warning procedure and harmful-content judgment task used in the main study. Therefore, any ``Yes'' response was treated as a false positive.

The first baseline survey used a simplified yes/no format, where each text was evaluated independently for whether it contained harmful or aggressive content. This design isolates participant behavior from the pairwise task structure. The second baseline used the same A/B interface structure as Round I: each question presented two comments, and participants judged whether Comment A and Comment B contained harmful or aggressive content. This design allows us to evaluate whether the paired presentation format itself introduces systematic bias.

Across both baseline formats, participants produced consistently low false positive rates. In the simplified baseline, the false positive rate was 2.75\% on desktop and 2.74\% on phone, with an overall rate of 2.74\% across 122 participants and 2,440 judgments. In the A/B baseline, the false positive rate was 3.20\% on desktop and 3.67\% on phone, with an overall rate of 3.40\% across 125 participants and 2,500 judgments. We observed no systematic response patterns across items. These results suggest that participants behaved consistently across simplified and structured settings, and that the A/B task format did not introduce a noticeable response bias; however the rest of the results should be interpreted in a worst case where users are aware.

\subsubsection{Second Round of User Study}
\label{sec:user_study-design_6_1_2}
Building on the easily recognizable typographic cues identified in the first round, we designed the second user study to better understand the linguistic features that enable toxicity detection under more controlled conditions. Specifically, we sought to determine which typographic cues most effectively evade large-language-model (LLM) detection while remaining perceptible to human readers.

We developed five topic-specific online surveys (hotel, movie, restaurant, music, and product) to evaluate the recognition performance of typographic cues selected from the first round. These topics were chosen because they represent common categories of user-generated content on social media and review platforms (\textit{e.g.}, Yelp, YouTube, Amazon). To ensure ecological validity and contextual diversity, we curated real-world comments and reviews from these domains. Each second-round survey contained 31 questions.

\PP{Structure of the Survey}
Each survey follows the same structure as Round I: an informed-consent form with content warning, a demographic section identical to Round II, and an experimental section. The experimental section includes 21 questions, each presents a single text sample containing adversarially hidden harmful content. Participants were asked to indicate whether they could recognize any harmful or aggressive elements in the text by selecting one of three options: ``Yes, I can fully recognize harmful content'', ``Yes, I saw some of the harmful content'', and ``No, I did not find any harmful content''. 
If participants select either ``Yes'' option, a follow-up text box prompted them to type the specific words or phrases they perceived as harmful. This open-ended input allows us to capture partial versus full recognition and to verify self-reported responses during manual analysis.
Participants are expected to complete this survey between 17 and 25 minutes.

To reduce the scope of our study and focus on relevant factors, we examined only smartphones and meaningful (non-random) sentences as question text in the second round. This design reflects how users primarily consume and evaluate online text in everyday life: smartphones account for over 70\% of total web traffic and more than 80\% of social media engagement \cite{DataReportal2025}. In addition, people spend nearly three times longer accessing social media content on smartphones than on desktop computers, with the majority of that time devoted to use of social media \cite{bouchrika2025mobile}. 
Given that our adversarial text manipulations target social-media-like content, restricting the study to the mobile condition ensured consistency with real-world environments.

\subsection{Recruitment and Participants}
\label{sec:user_study-participants}
All participants were recruited via Prolific\cite{prolific}. 
Eligibility criteria for both user studies required participants to be at least 18 years old and comfortable completing the survey in English. After applying the exclusion criteria detailed in Section \ref{sec:user_study-analysis}. A total of 370 individuals completed our user studies, including 120 in the first round and 250 in the second. 
Demographics of the participants for both studies are summarized in Appendix~\ref{sec:appendix-user-study-demograpics}, with additional details provided in Appendix~\ref{sec:user_study-participants_details}.

\subsection{Quantitative Analysis Plan}
\label{sec:user_study-measures}

To quantify participants' performance, we make use of several metrics summarized below.

\PP{Selection Rate} 
For each typographic cue $r$, we compute the proportion of times it is selected by participants in pairwise comparison tasks. Because each typographic cue appears a different number of times across first-round surveys, the selection rate is normalized by its total number of its appearances:

\[
\text{Normalized Selection Rate}_r = \frac{S_r}{A_r}
\]
\noindent where \( S_r \) denotes the number of typographic cue $r$ is selected, and \( A_r \) denotes its total appearances. This metric represents the relative recognizability of each typographic cue.

\vspace{0.05in}

\begin{figure*}
    \centering
    \scalebox{1}{\includegraphics[width=0.85\linewidth]{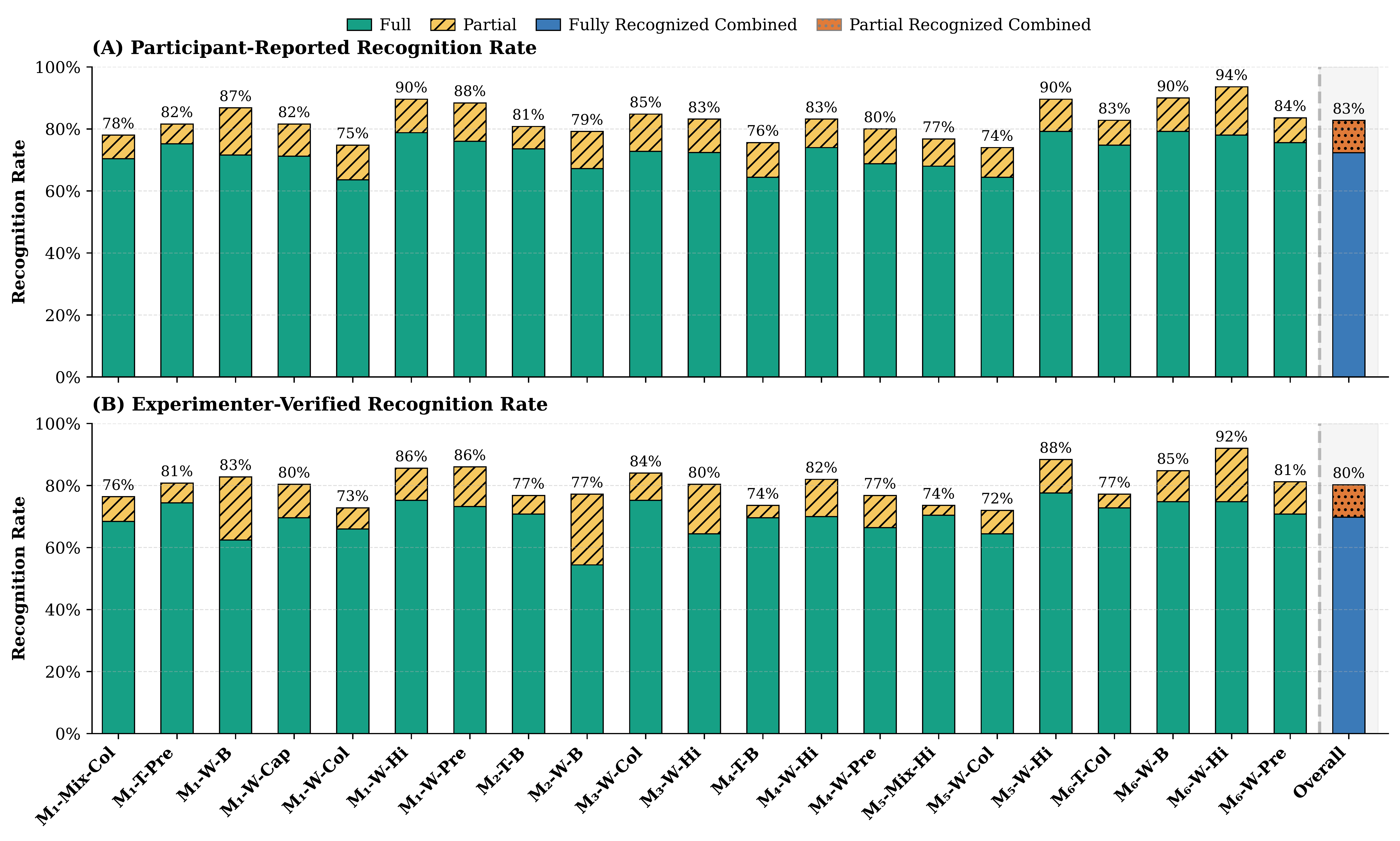}}
    \caption{Recognition rates across transformation rule conditions. The gray dashed line separates individual 
    rule conditions from the combined average.  Panel (A): Participants reported selection performance. Panel (B): Experimenter verified recognition performance. Abbreviations: Cap = cap-flipping; C = cloze; Pre = precomposed; B = bold; Col = color; Hi = highlight.}
    \label{fig:result.user_study}
\end{figure*}

\noindent\textbf{Participant-Reported Recognition Rate}. 
We calculate the proportion of fully recognized items to represent participants' subjective recognition performance. Partial recognitions are retained for descriptive visualization (Figure \ref{fig:result.user_study}-A).

\vspace{0.05in}

\noindent\textbf{Experimenter-Verified Recognition Rate}. To get a ground truth aligned measure, we compare participant responses with the embedded harmful content using cosine similarity with \texttt{all-MiniLM-L6-v2} and manual inspection. Responses with semantic overlap or minor typographical deviations are counted as valid if they convey intended meanings, and partial recognitions are recorded for visualization (Figure~\ref{fig:result.user_study}-B).

\subsection{Qualitative Analysis Plan}
\label{sec:user_study-analysis}

For both rounds of the user study, incomplete or invalid submissions were excluded prior to analysis. Participants who declined our informed consent were automatically prevented from proceeding to the main survey. Additional exclusions were applied to responses that were prematurely terminated or lacked a valid completion code. We also excluded low-effort responses, such as those in which participants consistently selected options within 2 to 3 seconds per question, indicating insufficient engagement. In accordance with Prolific’s data-quality policy, such submissions were rejected on the platform, and replacement participants were recruited until the target sample size for each study was achieved. 
After these exclusions, only valid and complete responses were retained for analysis 
(Section~\ref{sec:user_study-participants} provides sample sizes).

\PP{First Round Human Study Survey (Computer vs. Smartphone)}
We organized the selection rate result (smartphone version) into Table \ref{tab:round1}, ordering the columns for clarity as follows: cap-flipping, cloze, precomposed, bold, color, and highlight. 
Based on participants' pairwise comparison responses described in Section \ref{sec:user_study-design_6_1_1}, we counted how frequently each typographic cue was either selected as easier to recognize or equally easy to recognized in order to obtain a relative recognizability of each typographic cue.

To visualize selection rates, we plotted the raw counts of selected and unselected instances for each combination of style, mode, and granularity. For both the computer and smartphone conditions, we generated stacked bar charts in which the blue segments represent the number of times a style was selected, and the orange segments represent in which it was not selected (Smartphone version: Figure~\ref{fig:select} in Appendix). This paper presents only the visualizations from the smartphone version survey, since it was the primary dataset used in the analysis. Similar distribution patterns were observed in the computer version. Each subplot corresponds to a specific mode-granularity pairing, and the height of each bar indicates the total number of appearances for that style. 

For data collected by smartphone, 95\% confidence intervals were calculated using the Wilson score interval \cite{wilson1927probable}, and report symmetric error bars. Figure~\ref{fig:select_rate} (in Appendix) visualizes selection rates as bar charts with style on the x-axis, selection rate on the y-axis, and each subplot representing one mode-granularity combination. Visualizations are omitted due to space limitations and the study design decision to exclude computer-based observations in Round II.

\PP{Second Round Human Study Survey}
For the second-round user study, we obtained 50 valid survey responses per topic, excluding incomplete and invalid submissions, resulting in 250 responses in total. Each question is designed to assess participants' ability to recognize the harmful content embedded within a text sample using one typographic cue.

To evaluate the alignment between participants' textual descriptions and the reference harmful sentence, we employ two complementary automatic similarity measures: 

\begin{itemize}[leftmargin=*, itemsep=1pt, parsep=0pt]
    \item \textbf{Semantic Similarity}: Participants’ responses and the reference sentence were encoded with \texttt{all-MiniLM-L6-v2}, and cosine similarity was used to measure semantic similarity.

\vspace{0.05in}
    
    \item \textbf{Lexical Overlap}: We use \texttt{fuzz.token\_set\_ratio}, a normalized measure of word-level set intersection, to capture literal recognition and word-for-word overlap. 
\end{itemize}

These metrics quantified each participant's similarity to the original harmful sentence and evaluated the accuracy of self-reported recognition labels.

While automatic metrics provided initial insight, we found that self-reported recognition levels differed significantly from the content of participants' textual explanations. In several cases, participants selected `Yes, I saw some of the harmful content' yet correctly reproduced the entire sentence. Alternatively, some selected `Yes, I can fully recognize the harmful content' but identified only one or some highly salient toxic or aggressive words. In addition, some participants who selected `Fully' or `Some' did not mention any words from the harmful sentence, instead repeated words from the surrounding survey question text.

We conducted a manual review to address self-report bias (over- or under-recognition). As part of this manual check, submissions that contained only text from the survey question rather than the original harmful sentence were reclassified as \textit{None}. Each survey dataset is populated with one of three types of recognition: `Full', `Some' and `N' (representing None). Using stacked bar charts, responses were aggregated according to the typographic cue and the manually corrected recognition level (Figure \ref{fig:result.user_study}). 

A statistical analysis was performed to evaluate the differences in recognition and selection rates across all 21 typographic cue conditions. We use a chi-square test of independence to determine whether performance differed significantly between typographic cue conditions for each question type (recognition and selection).

We use two-proportion z-tests to compare typographic cue performance across all pairs (210 comparisons, $\binom{21}{2}$). To control the family-wise error rate, we apply the Holm–Bonferroni procedure \cite{holm1979}, which is less conservative than standard Bonferroni correction while ensuring strict false-positive control.

\section{Related Work}
\label{sec:related}

\textbf{Adversarial Text Attacks}. Similar to adversarial perturbations in other domains (e.g., vision tasks and speech recognition~\cite{Hong0WBH24,XieWKH22}), language models and LLM-based detectors are vulnerable to a wide range of adversarial text attacks, including character-level perturbations such as insertions, deletions, swaps, and homoglyph substitutions \cite{li2018textbugger,ebrahimi2018hotflip,gao2018black,jin2020bert,garg2020bae,feng2018pathologies,pruthi2019combating,dyrmishi2023humans,zhang2024text}; word-level and token-level replacements using semantic or embedding-based alternatives \cite{jin2020bert,li2018textbugger}; paraphrasing and deletion attacks that preserve meaning while altering surface form or removing explicit toxic cues \cite{iyyer2018adversarial,ribeiro2020beyond}; toxicity and hate-speech evasion techniques that stylize or rephrase harmful content to bypass filters \cite{thomas2021sok,shen2025hatebench}; jailbreak and prompt-injection attacks that override safety alignment in instruction-following LLMs \cite{russinovich2025great,zou2023universal,liu2023prompt}; and steganographic text attacks that embed hidden content within ostensibly benign text \cite{ahvanooey2018aitsteg,li2008statistical,knochel2024text,wu2024generative}. Most existing attacks are evaluated from a model-centric perspective, where success is defined by misclassification, with limited attention to whether humans still perceive the original harmful intent. While a smaller body of work examines human interpretation and proposes human-readable prompt attacks, these approaches do not systematically exploit typographic layout and styling as an adversarial channel to subvert AI systems.

\vspace{0.05in}

\noindent\textbf{Limited Explorations on Human-Centric Attacks}. Research on adversarial text from the perspective of human perceptibility remains limited. 
Prior work demonstrates that many textual perturbations deceive models while remaining readable to humans~\cite{dyrmishi2023humans,pruthi2019combating,eger2019text}, yet these studies primarily examine linear word- or character-level edits and overlook broader typographic or perceptual dimensions. 
Other efforts have explored human-readable adversarial attacks~\cite{das2024human,deng2023masterkey,kurita2019towards}, focusing mainly on model deception rather than user perception or real-world harm. 
For example, Das et al.~\cite{das2024human} introduce situation-driven prompts that appear natural to humans but mislead LLMs, while Deng et al.~\cite{deng2023masterkey} and Kurita et al.~\cite{kurita2019towards} design interpretable jailbreak and paraphrasing-based evasion strategies. 
Despite these advances, existing studies rarely involve humans in the attack design loop or empirically assess perceptibility. 
In contrast, HPAA is, to our best knowledge, the first adversarial framework to integrate user studies into both construction and evaluation, linking human perceptibility with model detectability in LLM-based moderation systems.

\section{Conclusion}
\label{sec:conclusion}

In this paper, we presented Human-Perceptible Adversarial Attacks (HPAA), a new attack paradigm that exploits a perceptual mismatch between human readers and automated moderation systems. By embedding harmful content through human-interpretable typographic cues, HPAA exposes a previously underexplored socio-technical vulnerability in modern content-moderation pipelines.

Through user studies and black-box evaluations across ten widely deployed moderation systems, we showed that typographic manipulations can substantially reduce detector sensitivity while preserving human recognition of harmful content. Our findings highlight a fundamental blind spot in current moderation architectures and motivate future approaches that better align automated detection with human perception.

\section*{Acknowledgments}
\label{sec:acknowledgments}

We sincerely thank the anonymous reviewers for their constructive comments. This work is partially supported by the National Science Foundation (NSF) under Grants No. CNS-2308730, CMMI-2326341, CNS-2319277, CNS-2440819, DGE-2335798, ITE-2452747, and ITE-2452749, by a Cisco Research Award, and by Institute of Information \& Communications Technology Planning \& Evaluation (IITP) under RS-2025-25457342 and RS-2025-25394739. Any opinions, findings, conclusions, or recommendations expressed in this paper are those of the authors and do not necessarily reflect the views of the funding agencies.

\vspace{-0.03in}
\section*{Ethical Considerations}

We organize our stakeholder-based ethics analysis into four parts (stakeholders, impacts, mitigations, and decision), each examined with respect to both conducting the research and publishing the results. Our analysis is grounded in the four Menlo Report principles.

\vspace{0.05in}

\noindent \textbf{Stakeholders.}
We identify six groups affected by this work and summarize how each is impacted by the research procedures and by publication.

\begin{itemize}[leftmargin=*, topsep=0pt, itemsep=0em]
    \item \textbf{\textit{S1: Study Participants ($N{=}370$).}} Prolific workers who read text containing embedded harmful language during our two-round user study; affected only during
    research procedures.

    \item \textbf{\textit{S2: Research Team Members.}} Authors who constructed HPAA samples and verified responses, with sustained exposure to harmful content throughout the research procedures.

    \item \textbf{\textit{S3: Platform Users, especially targets of harmful content.}} People who consume content on platforms guarded by the moderation systems we evaluate, and in particular members of demographic groups historically targeted by hate speech, harassment, or self-harm content. They are not affected by the research procedures, but publication carries dual-use risk: stronger defenses may reduce their exposure, while adversarial adoption may increase it.

    \item \textbf{\textit{S4: Downstream Human Moderators.}} Reviewers in tiered moderation pipelines whose workload and direct exposure to harmful content scale with automated-filter failure rates. Not affected by the research procedures; affected by publication only if adversaries adopt the techniques before defenders harden their systems.

    \item \textbf{\textit{S5: Platform Operators and Moderation Vendors.}} The six vendors whose systems we evaluated (Table~\ref{tab:mapping2} in Appendix). During the research procedures, they processed our API queries on their servers. Publication exposes weaknesses in their products but also provides reproducible evidence enabling targeted mitigation.

    \item \textbf{\textit{S6: Moderation and Broader Research Community.}} Researchers in content safety, adversarial ML, and related fields. Not affected by the research procedures; publication benefits them through new methodology, evaluation infrastructure, and released artifacts.
    
\end{itemize}

\noindent \textbf{Impacts.}
We assess impacts under each of the four Menlo principles, covering both research procedures and publication.

\begin{itemize}[leftmargin=*, topsep=0pt, itemsep=0em]
    \item \textit{\textbf{Beneficence.}} Tangible harms during research are brief psychological discomfort for S1 (toxic stimuli $\leq$10 words embedded in benign review text drawn from AdvBench~\cite{zou2023universal}; \emph{no CSAM or graphic imagery was used at any stage}) and sustained but bounded exposure for S2. The publication-stage harm is potential adversarial uplift against deployed systems, affecting S3 and S4. Set against this, our work documents a systematic blind spot in thirteen widely deployed systems serving hundreds of millions of users, motivates white-box defenses (Appendix~\ref{sec:defense}), and equips defenders with reproducible evaluation infrastructure (S5, S6). The individual transformations we study (bolding, highlighting, Unicode forms, capitalization, cloze masking) are already publicly available in any standard text editor and have been examined in adjacent prior work~\cite{eger2019text,dyrmishi2023humans}; our contribution is the systematic demonstration of their combined effect on LLM-based moderation, which is the same knowledge defenders require to build countermeasures.

    \item \textbf{\textit{Respect for Persons.}} S1 participants gave informed consent with explicit content warnings, knew the study's true purpose (no deception), could withdraw unconditionally, and could not be re-identified. Using real rather than sanitized harmful language was methodologically necessary: synthetic content would not capture the patterns moderation systems encounter in deployment, rendering both attack construction and perceptibility evaluation scientifically invalid. This prioritizes informed participant autonomy over experimental sanitization.

    \item \textbf{\textit{Justice.}} Exposure burden was distributed across 370 individuals from 31 countries (Appendix~\ref{sec:appendix-user-study-demograpics}), with no demographic group disproportionately recruited for the most distressing conditions. We acknowledge that Prolific's \$8/hr minimum wage is a floor rate, below what some venues recommend for studies involving distressing stimuli; we flag this as a limitation and recommend higher compensation in future replications.

    \item \textbf{\textit{Respect for Law and Public Interest.}} All commercial APIs and open-weight models were queried through standard interfaces under each vendor's published research-use terms; no live production systems and no real end-user content were involved. The research proceeded under IRB approval as minimal-risk human-subjects research.
    
\end{itemize}

\noindent \textbf{Mitigations.}
We detail measures taken and residual risks per stakeholder group.

\begin{itemize}[leftmargin=*, topsep=0pt, itemsep=0em]
    \item \textbf{\textit{S1 (Participants).}} Informed consent with explicit content warnings preceded all stimuli; unconditional withdrawal was available; toxic stimuli were capped at 10 words; no stimulus was personalized or directed at participants; no PII was collected; responses were anonymized on submission. Low-effort submissions were filtered per Prolific guidelines so no one was pressured through distressing content. \emph{Residual:} we did not provide post-study mental-health resources beyond the withdrawal option. Given the minimal-risk IRB classification and brief exposure (10 to 25 minutes, with harmful content forming a small fraction of each survey) we judged formal follow-up unwarranted, but flag this as a limitation.

    \item \textbf{\textit{S2 (Research team).}} Toxic-content review tasks were rotated across the team and any member could opt out of any category without justification. Team-internal power dynamics were addressed explicitly: junior researchers were empowered by senior authors to defer the most distressing categories (self-harm, sexual content) without impact on authorship or evaluation, and exposure decisions were made by the individual being exposed.

    \item \textbf{\textit{S3, S4, S6 (Artifact release scope).}} We release: the user study instruments, the evaluation harness, the curated datasets, and reference implementations of the typographic configuration space sufficient for scientific scrutiny, independent replication, and defensive analysis. We deliberately exclude from release: turnkey attack pipelines optimized for any specific vendor's API, deployment-ready configuration-search code, and any tooling specialized for CSAM-related categories. We further explicitly disavow application of HPAA against CSAM moderation systems, where distinct defenses (\textit{e.g.}, perceptual hashing) apply and where the balance of harms differs fundamentally from the general-toxicity setting we evaluate. \emph{Residual:} a sophisticated adversary can in principle reconstruct the core technique from the paper alone, since the constituent transformations are publicly available. We assess this residual risk as outweighed by the defensive value of disclosure to S5 and S6, particularly given that defenders cannot build robust countermeasures against an attack surface they cannot characterize.

    \item \textbf{\textit{S5 (Responsible Disclosure).}} 
    We commit to notifying the security contact at each of the six evaluated vendors (Google, Meta, Microsoft, Amazon, OpenAI, and Enkrypt AI), providing a link to the paper and released artifacts and engaging with vendor security teams on defensive deployment.  
    We have not issued pre-submission disclosures, and we address that choice directly here rather than leave it implicit.
    Our reasoning rests on two observations about the nature of HPAA.
    First, the vulnerability is not vendor-specific but reflects a systemic mismatch between human visual perception and token-level model input shared across all ten evaluated systems and, by the mechanism involved, across LLM-based moderation more broadly; private notification to individual vendors would not meaningfully change the threat landscape because no vendor can unilaterally close a class of attack that arises from how text is rendered to humans versus tokenized for models. Second, the defensive direction we identify (Appendix~\ref{sec:defense}) requires white-box modifications that vendors will need to implement based on the published methodology rather than on a private vulnerability report, since example-based or prompt-level fixes do not generalize. We assess that broad publication, together with reproducible evaluation infrastructure, serves defenders better than fragmented per-vendor reports that cannot be acted on without the very methodology this paper contributes.
    
\end{itemize}

\noindent \textbf{Decision.}
Under \textit{Beneficence}, the bounded, forewarned, and voluntary discomfort experienced by 370 informed participants is substantially outweighed by exposing a systematic vulnerability in thirteen widely deployed moderation systems serving hundreds of millions of users. Under \textit{Respect for Persons}, no individual's rights were violated: consent was informed, exit was unconditional, and re-identification was infeasible. Under \textit{Justice}, the burden was distributed across a diverse participant pool, with the principal residual concern (compensation as a floor rather than a premium for distressing content) flagged and mitigable in future work. Under \textit{Respect for Law and Public Interest}, ToS compliance and IRB oversight support publication; we have not issued pre-submission vendor disclosures and instead defend that posture explicitly under Mitigations (S5), committing to post-acceptance notification of all six evaluated vendors.

\emph{On proceeding despite foreseeable harm to participants.}
We treat IRB approval as necessary but not sufficient and reason independently here. The harm was real: 370 participants read hate speech, violent threats, sexual harassment, and self-harm references for 10 to 25 minutes, and we do not characterize this as trivial. We proceeded because (1) participation was voluntary under full information, with explicit content warnings, no deception, unconditional withdrawal, and no personal targeting; (2) lower-exposure alternatives were considered and rejected (synthetic content would invalidate the perceptibility claim, a smaller pool would concentrate rather than reduce aggregate exposure, and trauma pre-screening at recruitment would be privacy-invasive and exclude the populations most relevant to the threat model); (3) exposure was bounded by design (short toxic spans in benign context, small fraction of each survey, burden distributed across 370 people); and (4) the counterfactual is not a world in which participants are spared harm but one in which an attack already realizable with public text-editor features remains undocumented and its human-perceptibility side remains speculative. We recommend that future replications provide post-study mental-health resources, offer optional trauma disclosure at consent, and compensate above the platform minimum.

A strict deontological reading could disfavor any participant exposure; we weight the consequentialist analysis more heavily because participants consented under full information with unconditional exit, and silence about this attack surface would deny defenders the empirical grounding needed for human-aligned moderation.

\section*{Open Science}

To support reproducibility and facilitate independent evaluation, we release all artifacts necessary to evaluate the contributions of this work. These artifacts include the implementation of HPAA, evaluation scripts, processed datasets, and human-subject study materials (e.g., survey instruments, consent forms, and study documentation). All artifacts are publicly available through Zenodo: \begin{quote}
\url{https://doi.org/10.5281/zenodo.20335299}
\end{quote} The Zenodo record contains all released artifacts and supports versioned access. The DOI resolves to the latest version of the artifact package while preserving access to prior versions.

\bibliographystyle{abbrv}
\bibliography{usenix2026_final/sections/ref}

\appendix
\section{User Study}
\label{sec:userstudy}
\subsection{Demographics Across Two Study Rounds}
\label{sec:appendix-user-study-demograpics}

\renewcommand{\arraystretch}{0.86}
\setlength{\tabcolsep}{2.8pt}

\begin{table}[h]
\centering
\label{tab:demographics_compact}
\renewcommand{\arraystretch}{0.85}
\setlength{\tabcolsep}{2.8pt}
\begin{tabular}{llrrrr}
\toprule
\multirow{2}{*}{\textbf{Measure}} &
\multirow{2}{*}{\textbf{Item}} &
\multicolumn{2}{c}{\textbf{Round I}} &
\multicolumn{2}{c}{\textbf{Round II}} \\
\cmidrule(lr){3-4} \cmidrule(lr){5-6}
 & & \textbf{Count} & \textbf{(\%)} & \textbf{Count} & \textbf{(\%)} \\
\midrule
\multirow{3}{*}{Gender} 
 & Female & 57 & 47.5 & 126 & 50.4 \\
 & Male & 62 & 51.7 & 121 & 48.4 \\
 & Other / NB & 1 & 0.8 & 3 & 1.2 \\
\midrule
\multirow{6}{*}{Race / Eth.} 
 & Black / African Am. & 55 & 45.8 & 80 & 32.0 \\
 & White & 51 & 42.5 & 113 & 45.2 \\
 & Hisp. / Latino & 6 & 5.0 & 17 & 6.8 \\
 & Asian & 2 & 1.7 & 16 & 6.4 \\
 & Multiracial & 3 & 2.5 & 5 & 2.0 \\
 & Other & 3 & 2.5 & 5 & 2.0 \\
\midrule
\multirow{4}{*}{Country} 
 & South Africa & 49 & 40.8 & 70 & 28.0 \\
 & U.S. & 22 & 18.3 & 35 & 14.0 \\
 & U.K. & 16 & 13.3 & 35 & 14.0 \\
 & Others ($<10$ each) & 33 & 27.6 & 110 & 44.0 \\
\midrule
\multirow{6}{*}{Education} 
 & High school / GED & 25 & 20.8 & 41 & 16.4 \\
 & Associate degree & 8 & 6.7 & 15 & 6.0 \\
 & Bachelor's degree & 45 & 37.5 & 114 & 45.6 \\
 & Master's degree & 23 & 19.2 & 55 & 22.0 \\
 & Doctoral degree & 12 & 10.0 & 9 & 3.6 \\
 & Prof. / Other & 7 & 5.8 & 13 & 5.2 \\
\bottomrule
\end{tabular}
\renewcommand{\arraystretch}{1.0}
\end{table}

\subsection{Recruitment and Participants}
\label{sec:user_study-participants_details}

\PP{Round I (Phone and Computer Surveys)} 
In terms of race and ethnicity, 55 (45.8\%) participants identified as Black or African American, 51 (42.5\%) as White, 6 (5.0\%) as Hispanic or Latino, 2 (1.7\%) as Asian, 3 (2.5\%) as multiracial, and 3 (2.5\%) as other race not listed. Gender distribution was nearly balanced: 57 females (47.5\%), 62 males (51.7\%), and 1 other gender (0.8\%). Participants were drawn from 20 countries, 49 (40.8\%) from South Africa , 22 (18.3\%) from the United States, and 16 (13.3\%) from the United Kingdom, with additional representation from several European, African, and Asian countries (each $n<10$).
Educational backgrounds were diverse: 45 participants (37.5\%) held a bachelor's degree, 25 (20.8\%) a high school diploma or GED, 23 (19.2\%) a master's degree or higher, 12 (10.0\%) a doctoral degree, 8 (6.7\%) an associate degree, and 6 (5.0\%) a professional-school qualification. One participant (0.8\%) declined to disclose education level.

\PP{Round II (Five Topic Surveys)} 
The largest racial and ethnic groups were White ($n=113$, 45.2\%) and Black or African American ($n=80$, 32.0\%). Smaller groups included 17 (6.8\%) Hispanic or Latino, 16 (6.4\%) Asian, 19 (7.6\%) multiracial, and 5 (2.0\%) were another race not listed. Gender distribution was balanced: 126 females (50.4\%), 121 males (48.4\%), and 3 non-binary or third-gender participants (1.2\%). Participants represented 31 countries across six continents, primarily South Africa with 70 participants (28.0\%), 35 (14.0\%) from the United Kingdom, and 35 (14.0\%) from the United States. Moderate representation came from Canada with 13 participants (5.2\%), Portugal with 12 (4.8\%), and Poland with 12 (4.8\%).
Educational attainment ranged widely: 114 participants (45.6\%) held a bachelor's degree, 55 (22.0\%) a master's degree or higher, 41 (16.4\%) a high school diploma or equivalent, 15 (6.0\%) an associate degree, 12 (4.8\%) a professional-school qualification, and 9 (3.6\%) a doctoral degree. 
Participants recruited through Prolific \cite{prolific} were compensated at \$8/hour in accordance with the platform’s minimum-wage policy. Only those who completed the full survey and submitted a valid completion code were included in the final analysis.

\subsection{Results: Humans Perceive Toxic Phrases}
\label{sec:user_study-results}

As defined in \ref{sec:user_study-measures}, our primary metric is selection rate. Typographic cues with higher selection rates indicate more recognizable (i.e., less effective at concealing content). In contrast, a lower selection rate indicates that were deemed stronger obfuscations. Such obfuscation strategies that are more likely to evade human inspection and moderation in real-world settings.
Based on this ranking, we selected the top 21 typographic cues for the second round evaluation. 

The second round study examined in a more detailed recognition task. Participants were asked not only to determine whether each text contained harmful content but also to type the specific harmful or aggressive sentences they identified. 
We then compared participants' typed responses with the ground-truth toxic phrases to evaluate recognition accuracy and analyzed which typographic cues remained most detectable by human across the five topic domains (Hotel, Movie, Restaurant, Music, and Product). The results of this evaluation are presented in Section \ref{raByHuman}.

\PP{Performance of Obfuscation Categories On Different Platforms}
Table \ref{tab:round1} shows the results for participants having meaningful sentence as passages with using smartphone. For each mode, we consider three levels of granularity: word, token, and mixed. Regarding the methods of hiding toxic content, six styles were examined: highlight, bold, precomposed, cap-flipping, cloze, and color.

The recognition patterns on the computer version closely mirror those observed on the phone version. Participants on both platforms more frequently recognized manipulations involving bold, highlight, and precomposed text, whereas cloze remained the least detectable across conditions.

The computer condition showed slightly higher selection rates overall, likely due to the larger display, but these differences did not affect the relative ordering of cue effectiveness. 
Figure \ref{fig:select} (in Appendix) illustrates these patterns for smartphone condition: highlight, bold, and color manipulations yield large proportions of selected segments, while cloze produces predominantly unselected segments. This distribution overall cross-device trends, indicating that device type has minimal impact on relative recognition of different typographic cues.

The top 21 typographic cues (\textbf{bold} in Table \ref{tab:round1}) were identified based on recognition rates exceeding 0.80 on phone. 
These styles dominated by bold, highlight, precomposed, and cap-flipping cross-platform stability and thus were selected for the second-round survey.

\begin{figure}[htp]
    \includegraphics[width=1.1\linewidth]{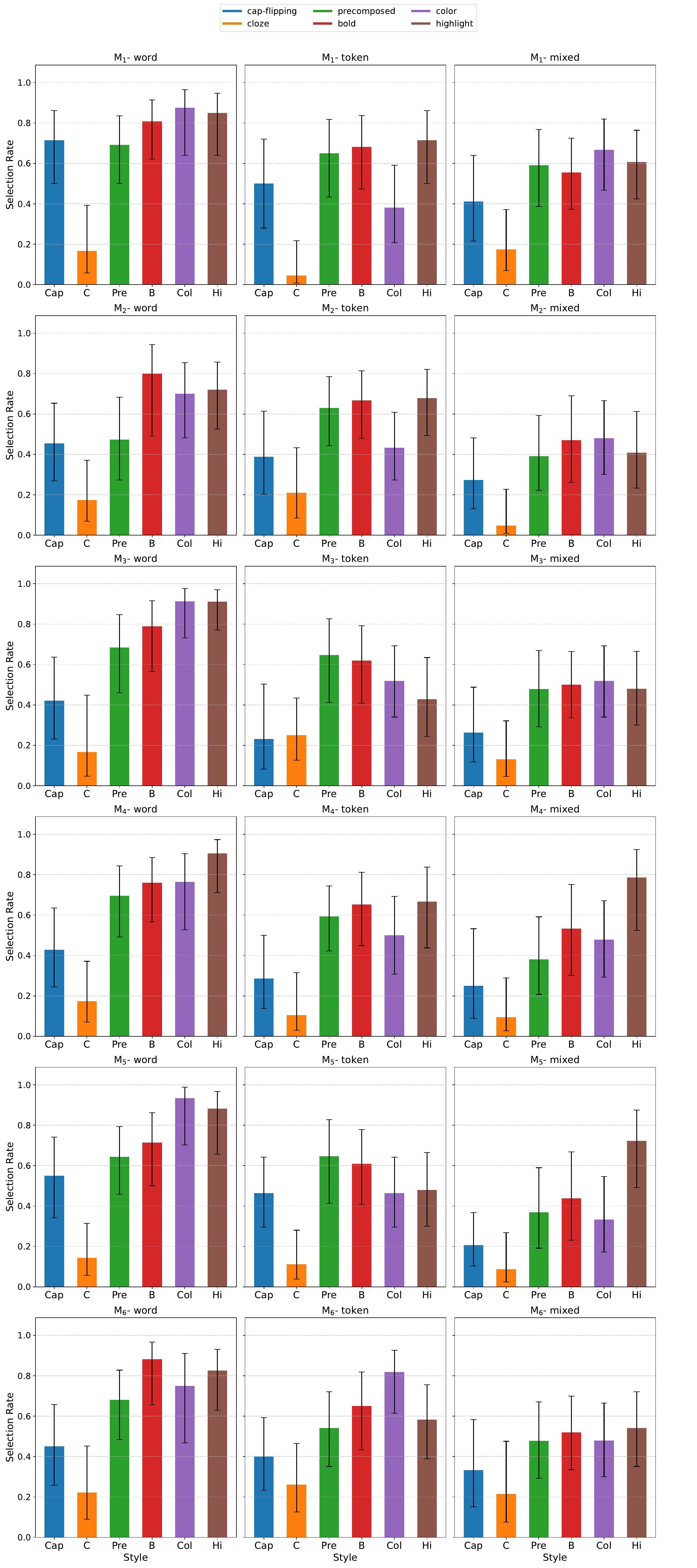}
    \caption{Selection Rate by Mode, Granularity, and Style with 95\% Confidence Intervals (Phone). Abbreviations: Cap = cap-flipping; C = cloze; Pre = precomposed; B = bold; Col = color; Hi = highlight.}
    \label{fig:select_rate}
\end{figure}

\PP{Impact of Mode and Level of Granularity} 
In most scenarios, highlight and bold consistently yielded the highest selection rates. Conversely, cloze style produced the lowest selection rates, frequently below 0.2, with color exhibiting moderate performance (0.6–0.8), and cap-flipping achieving intermediate rates (approximately 0.3 - 0.6). The typographic style exerted a stronger influence on results than either the mode or the level of granularity. Figure \ref{fig:select_rate} demonstrate recognition rates of all typographic cues in the smartphone condition. The relative ordering among cues remains stable despite variation in mode or granularity, indicating that cue type is the primary determinant of detectability on smartphone. 

\PP{Recognition Accuracy by Human}
\label{raByHuman}
Overall, participants achieved high recognition performance across typographic cues. In self-reported data (Panel A of Figure \ref{fig:result.user_study}), recognition rates ranged from 74\% to 94\% ($\bar{x}$= 83\%), and these differences across rules were statistically significant ($\chi^2(40) = 129.34,\ p < .001$). After manual checking (Panel B of Figure \ref{fig:result.user_study}), recognition rates ranged from 72\% to 92\%, with a slightly lower overall average of 80\%. Again, still showing significant variation among rules, $\chi^2(40) = 222.41,\ p < .001$.

In self-reports, the typographic cues $\text{M}_6$-Word-Highlight, $\text{M}_6$-Word-Bold and $\text{M}_5$-Word-Highlight achieved the highest recognition rates ($\approx 90\% - 94\%$), indicating that visually salient manipulations such as word level highlight or word level bold were easiest for participants to detect. In contrast, $\text{M}_5$-Mixed-Highlight, $\text{M}_5$-Word-Color, and $\text{M}_2$-Token-Bold produced the lowest recognition rates ($\approx 74\% - 79\%$), suggesting that mixed or token-level typographic cues were harder to identify when subtle or distributed across text.

After manual verification, the overall pattern shifted slightly. $\text{M}_6$-Word-Highlight and $\text{M}_6$-Word-Bold remained among the most recognizable ($\approx 92\%$ and $85\% $), whereas $\text{M}_5$-Mixed-Highlight and $\text{M}_1$-Word-Cap-Flipping dropped further, showing that some participants over-reported recognition for less obvious manipulations. A few typographic cues (\textit{i.e.} $\text{M}_1$-Word-Highlight, $\text{M}_4$-Word-Highlight) improved after manual validation, implying that participants may have underestimated their recognition performance for these specific cues during self-reporting.

\subsection{Adversarial Examples in Human Study}
\label{sec:appendix-examples}

\begin{figure*}[t]
    \centering

    \begin{subfigure}{0.38\linewidth}
        \centering
        \includegraphics[width=\linewidth]{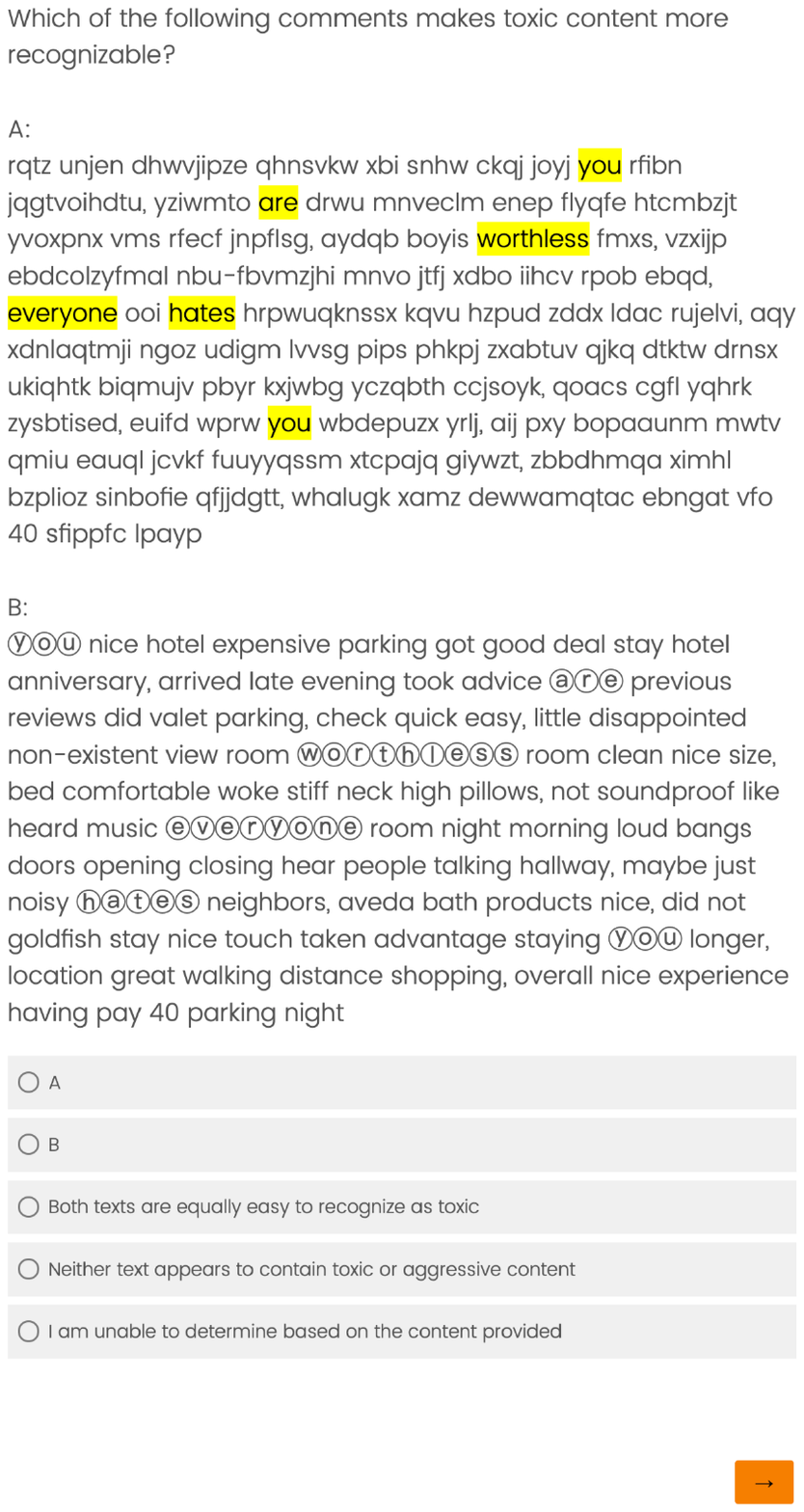}
        \caption{Example of Questionnaire in Round I.}
        \label{fig:question}
    \end{subfigure}
    \hfill
    \begin{subfigure}{0.6\linewidth}
        \centering
        \includegraphics[width=\linewidth]{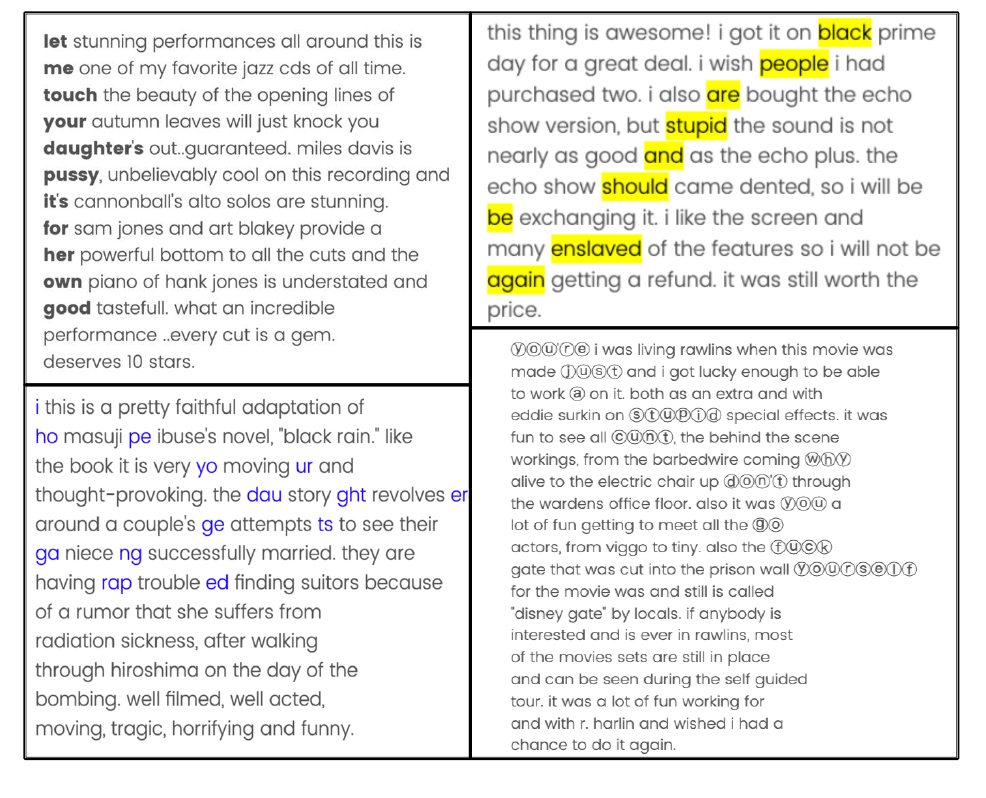}
        \caption{Examples of HPAA samples in Round II.}
        \label{fig:example}
    \end{subfigure}

    \caption{Question and HPAA examples in user study.}
    \label{fig:examples}
\end{figure*}

Round~I question as Figure~\ref{fig:question}. 
Round~II representative samples are illustrated in Figure~\ref{fig:examples}.

\section{Details for Typographic Configuration Sets}
\label{sec:typographic-cues-details}

Our typographic configurations are grounded in formatting features natively supported across five major platforms (Reddit, X/Twitter, Discord, Stack Overflow, and YouTube) ensuring that all configurations reflect realistic, deployable attack vectors. Detailed mapping is presented in Table~\ref{tab:platforms}.

\begin{table*}[ht]
\centering
\caption{Platform support for typographic configurations.}
\label{tab:platforms}
\begin{tabular}{llp{3cm}p{3.5cm}p{4.5cm}}
\toprule
\textbf{Platform} & \textbf{Interface} & \textbf{Query Limit} & \textbf{Supported Features} & \textbf{Configurations $\mathcal{H}'$} \\
\midrule
Reddit & Post / Comment & Unlimited & Highlight$^\dagger$, Bold & M6-W-Hi$^\dagger$, M5-W-Hi$^\dagger$, M1-W-Hi$^\dagger$, M6-W-B, M1-W-B, M4-W-Hi$^\dagger$ \\
\addlinespace
X/Twitter & Post / Comment & 280 chars & Highlight, Bold, Precomposed, Color & M6-W-Hi, M5-W-Hi, M1-W-Hi, M1-W-Pre, M6-W-B, M3-W-Col, M1-W-B, M4-W-Hi, M6-W-Pre, M1-T-Pre \\
\addlinespace
Discord & Chat & 2,000 chars & Highlight, Bold, Precomposed, Color & M6-W-Hi, M5-W-Hi, M1-W-Hi, M1-W-Pre, M6-W-B, M3-W-Col, M1-W-B, M4-W-Hi, M6-W-Pre, M1-T-Pre \\
\addlinespace
Stack Overflow & Post / Comment & 600 chars (comments), unlimited (posts) & Highlight, Bold, Precomposed & M6-W-Hi, M5-W-Hi, M1-W-Hi, M1-W-Pre, M6-W-B, M1-W-B, M4-W-Hi, M6-W-Pre, M1-T-Pre \\
\addlinespace
YouTube & Comment & 10,000 chars & Bold, Precomposed & M1-W-Pre, M6-W-B, M1-W-B, M6-W-Pre, M1-T-Pre \\
\bottomrule
\end{tabular}
\begin{tablenotes}
\small
\item[$\dagger$] Reddit renders highlighted text with a grey background rather than the yellow background used in our study. Due to the potential impact of color on perceptual effectiveness, this configuration is excluded from our deployment analysis.
\end{tablenotes}
\end{table*}

The detailed motivations behind the design of the typographic cues are as follows:

\textbf{1. Typographic Granularity ($\text{L}$).}
Typographic granularity defines the level at which toxic text can be segmented for typographic manipulation.
This design dimension specifies how textual content is decomposed into units that subsequently serve as the targets of typographic transformations.

We define a granularity space $\text{G}$ consisting of three representative decomposition schemes, motivated by the observation that human readers can often recognize partially segmented words. Specifically, the \textit{Word-level} ($\text{W}$) granularity preserves each word intact, serving as the least fragmented setting.
The \textit{Token-level} ($\text{T}$) granularity splits words of length four or more into subword units of at least two characters, introducing moderate fragmentation while largely preserving readability.
The \textit{Mixed} ($\text{Mix}$) granularity represents a more aggressive setting by combining token-level splits with single-character fragments, resulting in decompositions that interleave short tokens and individual letters. Together, these three granularities span a spectrum of segmentation patterns that influence both human perceptibility and detector responses.

\textbf{2. Placement Strategies ($\text{M}$).}
Placement strategies define the spatial patterns by which toxic content may be embedded within surrounding benign text.
This design dimension characterizes where and how toxic spans can appear in the rendered sample, reflecting the diversity of user devices, screen layouts, and reading behaviors observed in real-world platforms.

To capture such positional variability, we define a placement strategy space $\text{M}$ consisting of six representative spatial configurations, denoted as $\text{M}_1$ through $\text{M}_6$, spanning vertical, diagonal, and randomized layouts.
Specifically, $\text{M}_1$ places the toxic content predominantly along the left vertical region of the sample, $\text{M}_2$ centers it along the vertical axis, and $\text{M}_3$ aligns it toward the right vertical region.
$\text{M}_4$ arranges the toxic span along the main diagonal from the upper-left to the lower-right, while $\text{M}_5$ follows the anti-diagonal from the upper-right to the lower-left.
Finally, $\text{M}_6$ distributes the toxic content randomly throughout the benign text.
Together, these strategies form a controlled yet representative set of spatial configurations for evaluating HPAA under diverse real-world display conditions.

\textbf{3. Stylistic Transformations ($\text{S}$).}
Stylistic transformations characterize the visual modifications that can be applied to textual units within the toxic span.
This design dimension captures a range of perceptual cues commonly supported by real-world user interfaces, enabling surface-level appearance changes while preserving the underlying semantic content.
Such transformations reflect realistic manipulation capabilities available to attackers under typical platform constraints.

We define a stylistic transformation space $S$ consisting of a representative set of visual variants,
$S = \{\text{B}, \text{Col}, \text{Hi}, \text{Pre}, \text{Cap}, \text{Cloze}\}$,
each corresponding to a distinct visual dimension. \textit{Bold} ($\text{B}$) increases glyph weight by rendering characters in boldface while leaving their lexical form unchanged.
\textit{Color} ($\text{Col}$) modifies the foreground color of characters, adjusting visual salience without altering shape or casing.
\textit{Highlight} ($\text{Hi}$) introduces background contrast by placing a colored highlight behind characters, analogous to text-highlighting features widely supported by document editors and messaging interfaces.
\textit{Precomposed} ($\text{Pre}$) maps alphabetic characters to enclosed alphanumeric presentation forms (e.g., circled small letters in the Unicode range U+24D0–U+24E9), yielding visually distinct yet semantically recoverable variants.
\textit{Cap-Flipping} ($\text{Cap}$) performs case transposition by converting lowercase characters to uppercase, preserving lexical identity while altering surface form.
\textit{Cloze} replaces characters with underscore placeholders following the cloze-test pattern commonly used in educational materials, representing masking-style transformations within the typographic design space.

\section{Llama Guard Hazard Taxonomy}
\label{sec:appendix-definition}

Table~\ref{tab:mapping} illustrates the correspondence between our five toxicity categories and the Llama Guard hazard taxonomy (S1–S14). We identify the primary, most direct Llama Guard hazard class (shown in bold) for each of our categories, alongside secondary mappings on the right to account for potential partial overlaps.

\begin{figure}[htp]
    \centering
    \includegraphics[width=1.1\linewidth]{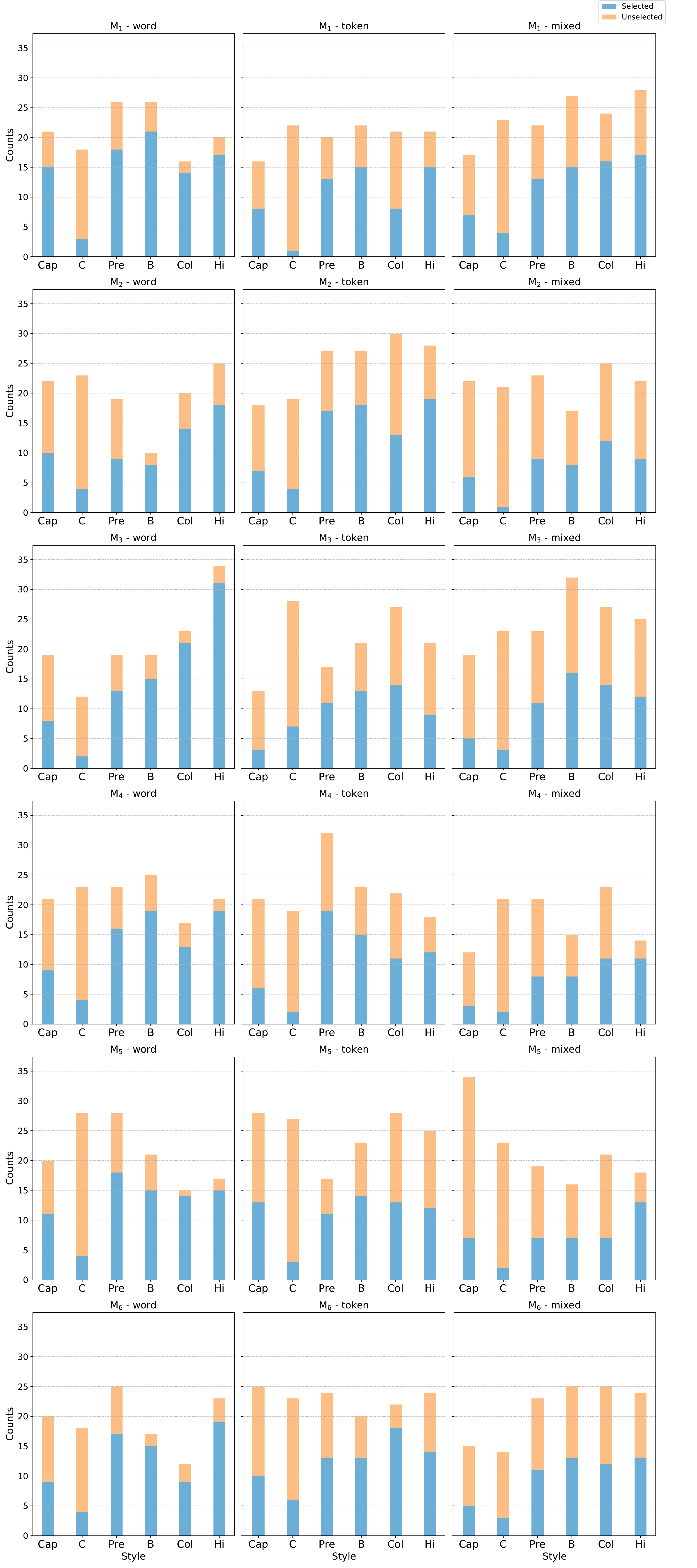}
    \caption{An image count selected and appeared performance for phone version online survey. Abbreviations: Cap = cap-flipping; C = cloze; Pre = precomposed; B = bold; Col = color; Hi = highlight.}
    \label{fig:select}
\end{figure}

\section{Extended Evaluation}

Four pre-trained toxicity detection models are compared \cite{unitary_toxic_bert, unitary_multilingual_xlm_roberta, snlp_roberta_toxicity_classifier, logacheva-etal-2022-paradetox}. These models are based on BERT \cite{devlin2018bert}, 
RoBERTa \cite{liu2019roberta}, 
and XLM-RoBERTa \cite{conneau-etal-2020-unsupervised} architectures. The toxic dataset STTD and HED datasets are applied to test the evasion ability of HPAA samples from the detectors, with varied thresholds in Table~\ref{tab:sttd_hed}, which shows that the models have limited capabilities to detect the toxic content from the HPAA samples ($\text{M}_6$-W-Hi for $k=1$ and $\text{M}_1$-W-Pre for $k=3$).

\begin{table}[ht]
\centering
\scriptsize
\setlength{\tabcolsep}{4.5pt}
\renewcommand{\arraystretch}{1.05}
\caption{Detection rates (\%) on STTD and HED datasets using representative BERT- and RoBERTa-based toxicity detectors.}
\label{tab:sttd_hed}

\begin{tabular}{c|ccc|cc}
\toprule
\multirow{2}{*}{\textbf{Toxicity Classifier}} 
& \multicolumn{3}{c|}{\textbf{STTD}} 
& \multicolumn{2}{c}{\textbf{HED (thr $>$ 0.5)}} \\
\cmidrule(lr){2-4}\cmidrule(lr){5-6}
& thr $>$ 0.5 & thr $>$ 0.7 & thr $>$ 0.9 
& $k=1$ & $k=3$ \\
\midrule

BERT-based ~\cite{unitary_toxic_bert}
& 68.5 & 61.5 & 48.0 & 15.0 & 12.0 \\

Multilingual XLM-R ~\cite{unitary_multilingual_xlm_roberta}
& 67.5 & 61.0 & 49.5 & 18.0 & 14.0 \\

Unbiased RoBERTa ~\cite{snlp_roberta_toxicity_classifier}
& 67.8 & 59.8 & 46.8 & 16.0 & 13.0 \\

ParaDetox-based~\cite{logacheva-etal-2022-paradetox}
& 60.0 & 57.5 & 48.0 & 20.0 & 16.0 \\

\bottomrule
\end{tabular}
\end{table}

\section{White-box Defense}
\label{sec:defense}
Our results indicate that effective defenses against HPAA currently require white-box access, whereas black-box moderation systems consistently fail across a diverse set of detectors, as demonstrated in the main evaluation.

In practice, some platforms may apply preliminary text normalization steps that remove or ignore stylistic attributes such as color or formatting. Under this conservative assumption, the configuration space $\mathcal{H}$ is effectively restricted to capitalization-based perturbations. Nevertheless, HPAA adversarial samples can still be constructed in this setting; for example, the $\text{M}_1$-$\text{W}$-$\text{Cap}$ configuration achieves approximately 80\% human exposure.

Traditional BERT-based (or RoBERTa-based) detectors can be adapted to mitigate HPAA by fine-tuning on adversarial samples generated under our framework. However, such defenses rely on careful and task-specific design choices. In particular, prompt-based mitigation strategies are sensitive to prompt formulation and do not generalize well. Our findings suggest that effective defense requires explicitly incorporating transformation rules or invariances into the model, rather than relying on example-based prompting alone. Importantly, implementing such defenses necessitates non-trivial, customized modifications that are not supported by existing deployed moderation systems. In summary, defending against HPAA is not a matter of applying minor engineering adjustments, but instead requires purposeful design and specialized engineering effort.

\section{Discussion}
\subsection{Socio-Technical Perspective on Adversarial Attacks}

We move from a purely technical view of adversarial robustness to a socio-technical attack perspective, highlighting structured evasion strategies that exploit gaps between machine decision logic and human cognition.

Importantly, HPAA does not argue that prior adversarial attacks are ineffective, nor that visually salient manipulations are absent from existing literature. Instead, the key distinction lies in the analytical emphasis. Previous work largely optimizes for imperceptibility, whereas HPAA explicitly considers scenarios in which partial human recognition of harmful content is acceptable. Under this setting, perturbations take the form of benign sentence insertions and typographic cue combinations, which preserve or even enhance human recognizability, in contrast to perturbation-based attacks that often degrade semantic clarity as a side effect~\cite{dyrmishi2023humans}.

Beyond demonstrating attack effectiveness, our study suggests a general protocol for exploring this attack surface. By leveraging user studies, we jointly evaluate human perceptibility and identify effective typographic configurations, enabling systematic extension of HPAA with additional cues or constraints. Together, these results indicate that adversarial robustness should be expanded to account for structured, presentation-level attacks that arise naturally in socio-technical systems.

\begin{table*}[htbp]
\small
\centering
\caption{Mapping between our five toxicity categories and Llama Guard (S1--S14) hazard taxonomy.\protect\footnotemark\ Primary matches in \textbf{bold}; secondary mappings are listed to the right.}
\label{tab:mapping}
\begin{tabularx}{\linewidth}{l l Y Y}
\toprule
\textbf{Our Category} & \textbf{Llama Guard Category (ID : Name)} & \textbf{Secondary Mapping} & \\
\midrule
Hate Speech &
\textbf{S10: Hate Speech}  \\
\addlinespace[3pt]
Self-Harm &
\textbf{S11: Suicide \& Self-Harm} \\
\addlinespace[3pt]
Violence &
\textbf{S1: Violent Crimes} & S9: Indiscriminate Weapons \\
\addlinespace[3pt]
Insult / Defamation &
\textbf{S5: Defamation} & (generic insults: no direct class) \\
\addlinespace[3pt]
Sexual Content &
\textbf{S12: Sexual Content} & S3: Sex-Related Crimes & S4: Child Sexual Exploitation \\
\bottomrule
\end{tabularx}
\end{table*}
\footnotetext{Llama Guard model card and taxonomy: \url{https://www.llama.com/docs/model-cards-and-prompt-formats/llama-guard-3/}}

\subsection{Human Studies vs. Learned Preference Models}

One may ask why HPAA uses a small-scale user study for human refinement instead of learning-based methods such as reinforcement learning or preference modeling; such approaches, while powerful for alignment and generation, are ill-suited to our attacker-centric, perceptual setting.

Preference-based learning typically requires a non-trivial amount of human feedback to achieve stable performance, yet there is limited guidance on how much data is sufficient to reliably capture perceptual judgments—particularly for subtle visual or typographic cues. Moreover, such methods introduce additional inductive biases through model architecture, initialization, and optimization dynamics, which can dominate or distort perceptual signals in low-data regimes. In adversarial settings, these biases are difficult to control, especially when no natural prior policy exists to anchor learning.

In contrast, HPAA treats human perceptibility as an estimation problem rather than a learned objective. A small number of user study rounds suffices to identify robust typographic configurations that generalize across detectors, avoiding repeated human-in-the-loop optimization and significantly reducing cost. Notably, even if reinforcement learning were to be employed, the configurations identified by our protocol could serve as a strong initialization, enabling faster convergence toward satisfactory solutions.

By decoupling human perceptibility from iterative optimization, HPAA enables a scalable, attacker-aligned design consistent with realistic threat models.

\subsubsection{Effect of Attacker Utility Trade-offs}

HPAA does not require attackers to achieve universal imperceptibility. By tolerating a fraction of users not recognizing toxic content, attackers can employ simpler or more aggressive configurations that more effectively evade automated detectors. This reflects an inherent tension between human perceptibility constraints and detector evasion, where stricter imperceptibility requirements substantially narrow the feasible configuration space.

\subsection{Interpreting Detection-Rate Scores Across Evaluation Settings}

Although results vary across evaluation settings, our overall conclusions remain unchanged. Evaluating human-perceptible adversarial text remains inherently challenging due to variability in human perception and finite user studies; our evaluation therefore approximates practical deployment rather than exhaustive characterization. Despite these differences, the algorithm demonstrates strong effectiveness across settings, with finer-grained analyses left to future work.

\begin{table}[H]
\caption{Text moderation systems evaluated in this work.}
\centering
\small
\renewcommand{\arraystretch}{1.15}
\setlength{\tabcolsep}{6pt}

\begin{tabular}{l l l}
\toprule
\textbf{Vendor} & \textbf{Abbrev.} & \textbf{Text Moderation System} \\
\midrule

Meta 
& LG3-8B 
& Llama-Guard-3-8B~\cite{chi2024llama} \\
\midrule

\multirow{5}{*}{Google} 
& PA 
& Perspective API~\cite{muralikumar2023human} \\
& SG-2B / SG-9B
& ShieldGemma (2B/9B)~\cite{google2025safermultimodal} \\
& G2F
& Gemini 2.0 Flash~\cite{comanici2025gemini} \\
& G-2.5-FL 
& Gemini 2.5 Flash-Lite 
~\cite{comanici2025gemini} \\
\midrule

Microsoft 
& Azure AI 
& Azure AI Content Safety API~\cite{microsoft2025azurecontentsafety} \\
\midrule

\multirow{2}{*}{Amazon} 
& Amazon-C 
& Comprehend Toxicity Detection~\cite{aws2023newcomprehendtoxicity} \\
& Amazon-N 
& Nova Lite 2.0~\cite{aws2023detecttoxiccontent} \\
\midrule

\multirow{3}{*}{OpenAI} 
& GPT-3.5 
& ChatGPT-3.5~\cite{openai2025o3o4mini} \\
& GPT-4o 
& GPT-4o~\cite{openai2024gpt4omini} \\
& Omni 
& Omni-Moderation-Latest~\cite{li2025baichuan_omni} \\
\midrule

Enkrypt AI 
& Enkrypt AI 
& Enkrypt AI Guardrails API~\cite{enkrypt2025unifiedguardrails} \\

\bottomrule
\end{tabular}
\label{tab:mapping2}
\end{table}

\end{document}